\documentclass{article}

\usepackage{PRIMEarxiv}

\usepackage[utf8]{inputenc} 
\usepackage[T1]{fontenc}    
\usepackage{hyperref}       
\usepackage{url}            
\usepackage{booktabs}       
\usepackage{amsfonts}       
\usepackage{nicefrac}       
\usepackage{microtype}      
\usepackage{lipsum}
\usepackage{graphicx}
\graphicspath{{media/}}     
\usepackage{tikz}
\usetikzlibrary{positioning} 
\usepackage{amsmath}
\usepackage{subcaption}
\newcommand{\upi}{\mathrm{i}}
\usepackage{multicol}
\usepackage{tabularray}
\usepackage{enumitem}

\title{Wave scattering at a rectangular junction of four waveguides
}

\author{
    Afnan A. Aldosri \\
    School of Information and Physical Sciences \\
    University of Newcastle, Newcastle \\
    \texttt{afnanabdullahm.aldosri@uon.edu.au} \\
    \\
    Michael H. Meylan \\
    School of Information and Physical Sciences \\
    University of Newcastle, Newcastle \\
    \texttt{mike.meylan@newcastle.edu.au} \\
    \\
    Ben Wilks \\
    School of Information and Physical Sciences \\
    University of Newcastle, Newcastle \\
    \texttt{ben.wilks@newcastle.edu.au}
}

\begin{document}
\maketitle

\begin{abstract}
We consider the scattering of linear waves in two dimensions by a rectangular region at the junction of four waveguides. A solution to the frequency domain problem is obtained by exploiting reflective symmetry to reduce the full problem to sub-problems defined on one quadrant of the junction. These sub-problems are solved using the eigenfunction matching method. The solution to the problem on the full region is then recovered from the solutions to the sub-problems, and a scattering matrix for the junction is presented. Finally, the solution in the time domain is constructed as a superposition of the frequency domain solutions and visualised for a range of incident pulses and waveguide geometries. 
\end{abstract}

\keywords{ Waveguides\and  Eigenfunction expansion \and Wave propagation}

\section{Introduction}
Waveguides have numerous applications across science and technology. They are critical components of electromagnetic and optical systems, acting as communication and sensing technologies \cite{riley2006mathematical}. They are also important in other wave-physics scenarios, such as acoustics and water waves \cite{mei2005theory}. In this latter topic, in which waveguides are more commonly referred to as channels, practical applications include wave flumes and towing tanks used in experimental research. In acoustics, sound propagation through the ducts of air-conditioning systems can also be modelled using waveguides as a framework \cite{Chapman2008, Munjal}.

The simplest problem to consider is how acoustic, water and electromagnetic waves diffract when encountering parallel half-planes. These investigations have been carried out under various boundary conditions, providing insight into wave behaviour in different contexts, for example, \cite{hassanMike2018, Lawrie201333, Girier2019, Lawrie2006}, \cite{Meylan2016, Hassanmike2009, Dalrymple1996} and \cite{Hameş2004, Roger22016}.

Various studies have investigated wave propagation in multi-furcated waveguides under different boundary conditions.  For example, bifurcated waveguides have been analyzed for their acoustic wave characteristics in discontinuous structures, considering the effects of soft and hard interfaces \cite{Afsar202343, Afsarbe2023}. Similarly, research on trifurcated waveguides has explored noise reduction techniques using different porous materials, focusing on their application in silencer modelling \cite{Afsar202444, hassan2016, hassan2014} and the behaviour of wave scattering in more complex structures, such as pentafurcated ducts, has been studied by \cite{hassan2017, hassanMah2017}.   

The study of wave scattering has extensively explored scenarios involving one or two step-down or step-up cascading discontinuities, addressing rigid boundary conditions \cite{Khalid2019, Nawaz2019, Afsar202343, Shung1971, Lewin1970, Afsar220024}, flexible boundary conditions \cite{Afzal2014, Afsar2021}, and mixed conditions combining both rigid and flexible \cite{Bilal2022, Satti2019, Boral2023}. In addition, several studies have focused on fluid-structure interactions at discontinuous interfaces between waveguides, examining how coupled waves scatter under these conditions \cite{Afzal2014, Afzal2016, Nawaz2013, Warren2002}. In addition, \cite{Satti2019} investigated acoustic wave scattering in cavities divided by parallel partitions, while \cite{Meylan2017} used symmetrical configurations to solve for wave behaviour in a resonator located within a conduit. In contrast, \cite{Christie2016, Ali2023} explored asymmetric solutions for wave scattering and validated their findings. 

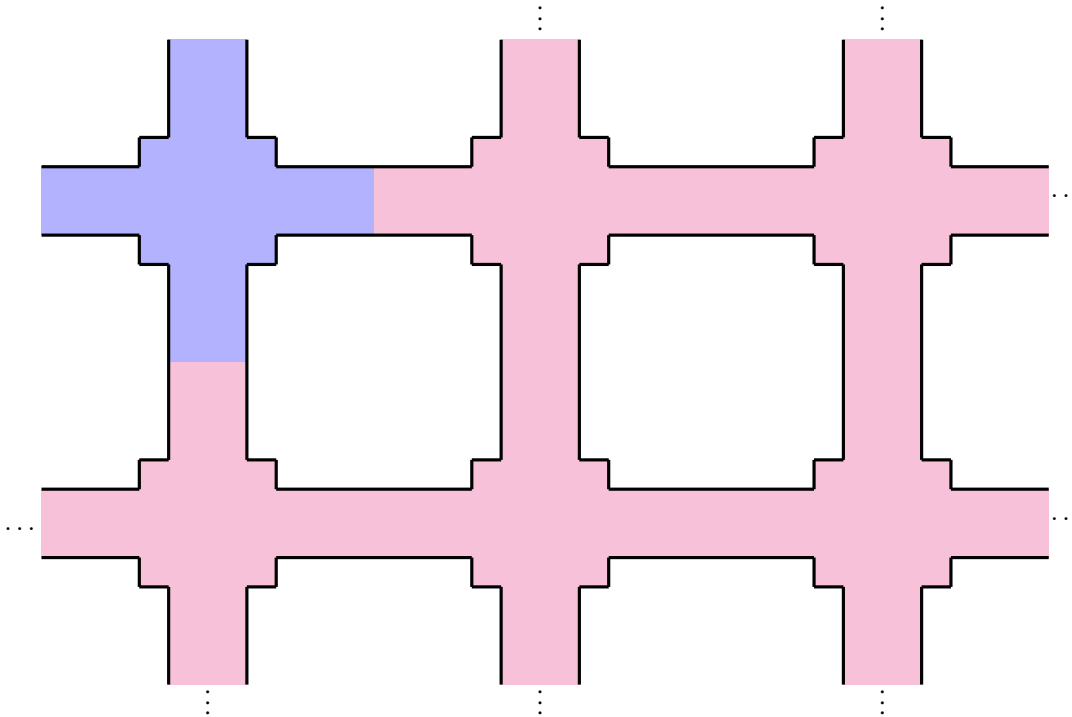
\begin{figure}[h]
\centering
 \begin{tikzpicture}[x=1.3cm,y=1.3cm][my triangle/.style={-{Triangle[width=\the\dimexpr1.8\pgflinewidth,length=\the\dimexpr0.8\pgflinewidth]}}]
\fill[blue!30!white] (-6.5,0) rectangle (-5.4,.7); 
\fill[blue!30!white] (-5.5,-.3) rectangle (-4.1,1); 
 \fill[blue!30!white] (-5.2,.9) rectangle (-4.4,2);
 \fill[blue!30!white] (-4.2,0) rectangle (-3.1,.7);
 \fill[blue!30!white] (-5.2,-1.3) rectangle (-4.4,-0.2);
\fill[magenta!30!white] (-3.1,0) rectangle (-2,.7);
\fill[magenta!30!white] (-5.2,-2) rectangle (-4.4,-4.6);
\fill[magenta!30!white] (-1.8,-2) rectangle (-1,-4.6);
\fill[magenta!30!white] (2.5,-2) rectangle (1.7,-4.6);
 \fill[magenta!30!white] (-.7,-.3) rectangle (-2.1,1);
 \fill[magenta!30!white] (2.8,-.3) rectangle (1.4,1);
\fill[magenta!30!white] (2.8,-2.3) rectangle (1.4,-3.6);
\fill[magenta!30!white] (-5.5,-2.3) rectangle (-4.1,-3.6);
\fill[magenta!30!white] (-.7,-2.3) rectangle (-2.1,-3.6);
\fill[magenta!30!white] (-5.2,-2.4) rectangle (-4.4,-1.3);
\fill[magenta!30!white] (-6.5,-3.3) rectangle (-5.4,-2.6);
 \fill[magenta!30!white] (-4.2,-3.3) rectangle (-2,-2.6);
\fill[magenta!30!white] (-.8,-3.3) rectangle (1.5,-2.6);
\fill[magenta!30!white] (3.8,-3.3) rectangle (2,-2.6);
\fill[magenta!30!white] (2.5,-2.4) rectangle (1.7,-.3);
\fill[magenta!30!white] (2.5,2) rectangle (1.7,.9);
\fill[magenta!30!white] (-1.8,2) rectangle (-1,.9);
\fill[magenta!30!white] (-1.8,-2.4) rectangle (-1,-.3);
\fill[magenta!30!white] (-.8,0) rectangle (1.5,.7);
\fill[magenta!30!white] (2.4,0) rectangle (3.8,.7);    
\usepgflibrary{arrows} 
\node[color =black, very thick]at(4,.4){$\dots$};
\node[color =black, very thick]at(4,-2.9){$\dots$};
\node[color =black, very thick]at(2.1,2.3){$\vdots$};
\node[color =black, very thick]at(-1.4,2.3){$\vdots$};
\node[color =black, very thick]at(-6.7,-3){$\dots$};
\node[color =black, very thick]at(-4.8,-4.7){$\vdots$};
\node[color =black, very thick]at(-1.4,-4.7){$\vdots$};
\node[color =black, very thick]at(2.1,-4.7){$\vdots$};
 \draw[color = black, very thick] (-5.2,1)--(-5.2,2);
\draw[color = black, very thick] (-4.4,1)--(-4.4,2);
\draw[color = black, very thick] (-5.2,1) -- (-5.5,1) node[left]{};
\draw[color = black, very thick] (-4.1,1) -- (-4.4,1) node[left]{};
\draw[color = black, very thick] (-5.5,1) -- (-5.5,.7); 
\draw[color = black, very thick] (-4.1,1) -- (-4.1,.7); 
\draw[color = black, very thick] (-5.5,.7) -- (-6.5,.7);
\draw[color = black, very thick] (-5.5,0) -- (-6.5,0);
\draw[color = black, very thick] (-4.1,.7) -- (-2.1,.7);
\draw[color = black, very thick] (-4.1,0) -- (-2.1,0);
\draw[color = black, very thick] (-2.1,1) -- (-2.1,.7); 
\draw[color = black, very thick] (-2.1,1) -- (-1.8,1); 
\draw[color = black, very thick] (-2.1,-.3) -- (-2.1,0); 
\draw[color = black, very thick] (-2.1,-.3) -- (-1.8,-.3); 
\draw[color = black, very thick] (-1.8,1) -- (-1.8,2);
\draw[color = black, very thick] (-1,1) -- (-1,2);
\draw[color = black, very thick] (-1,1) -- (-.7,1);
\draw[color = black, very thick] (-.7,1) -- (-.7,.7);
\draw[color = black, very thick] (-.7,.7) -- (1.4,.7);
\draw[color = black, very thick] (1.4,.7) -- (1.4,1);
\draw[color = black, very thick] (1.4,1) -- (1.7,1);
\draw[color = black, very thick] (1.7,1) -- (1.7,2);
\draw[color = black, very thick] (2.5,1) -- (2.5,2);
\draw[color = black, very thick] (2.5,1) -- (2.8,1);
\draw[color = black, very thick] (2.8,1) -- (2.8,.7);
\draw[color = black, very thick] (2.8,.7) -- (3.8,.7);
\draw[color = black, very thick] (2.8,0) -- (3.8,0);
\draw[color = black, very thick] (2.8,0) -- (2.8,-.3);
\draw[color = black, very thick] (2.8,-.3) -- (2.5,-.3);
\draw[color = black, very thick] (2.5,-.3) -- (2.5,-2.3);
\draw[color = black, very thick] (2.5,-2.3) -- (2.8,-2.3);
\draw[color = black, very thick] (2.8,-3.3) -- (2.8,-3.6);
\draw[color = black, very thick] (2.8,-2.6) -- (3.8,-2.6);
\draw[color = black, very thick] (1.7,-.3) -- (1.7,-2.3);
\draw[color = black, very thick] (1.4,0) -- (1.4,-.3);
\draw[color = black, very thick] (1.4,-.3) -- (1.7,-.3);
\draw[color = black, very thick] (1.4,0) -- (-.7,0);
\draw[color = black, very thick] (-.7,-.3) -- (-.7,0);
\draw[color = black, very thick] (-.7,-.3) -- (-1,-.3);
\draw[color = black, very thick] (2.8,-3.3) -- (3.8,-3.3);
\draw[color = black, very thick] (2.8,-2.3) -- (2.8,-2.6);
\draw[color = black, very thick] (2.8,-3.6) -- (2.5,-3.6);
\draw[color = black, very thick] (2.5,-3.6) -- (2.5,-4.6);
\draw[color = black, very thick] (1.7,-3.6) -- (1.7,-4.6);
\draw[color = black, very thick] (1.7,-3.6) -- (1.4,-3.6);
\draw[color = black, very thick] (1.4,-3.6) -- (1.4,-3.3);
\draw[color = black, very thick] (1.4,-2.6) -- (-0.7,-2.6);
\draw[color = black, very thick] (-0.7,-3.3) -- (-0.7,-3.6);
\draw[color = black, very thick] (-0.7,-2.3) -- (-1,-2.3);
\draw[color = black, very thick] (-1,-3.6) -- (-1,-4.6);
\draw[color = black, very thick] (-1.8,-3.6) -- (-1.8,-4.6);
\draw[color = black, very thick] (-2.1,-3.6) -- (-1.8,-3.6);
\draw[color = black, very thick] (-2.1,-3.6) -- (-2.1,-3.3);
\draw[color = black, very thick] (-2.1,-3.3) -- (-4.1,-3.3);
\draw[color = black, very thick] (-4.1,-3.3) -- (-4.1,-3.6);
\draw[color = black, very thick] (-4.1,-3.6) -- (-4.4,-3.6);
\draw[color = black, very thick] (-4.4,-3.6) -- (-4.4,-4.6);
\draw[color = black, very thick] (-5.2,-3.6) -- (-5.2,-4.6);
\draw[color = black, very thick] (-5.2,-3.6) -- (-5.5,-3.6);
\draw[color = black, very thick] (-5.5,-3.6) -- (-5.5,-3.3);
\draw[color = black, very thick] (-5.5,-3.3) -- (-6.5,-3.3);
\draw[color = black, very thick] (-1,-.3) -- (-1,-2.3);
\draw[color = black, very thick] (-1.8,-.3) -- (-1.8,-2.3);
\draw[color = black, very thick] (-1,-3.6) -- (-.7,-3.6);
\draw[color = black, very thick] (-.7,-2.3) -- (-.7,-2.6);
\draw[color = black, very thick] (-.7,-3.3) -- (1.4,-3.3);
\draw[color = black, very thick] (-2.1,-2.3) -- (-1.8,-2.3); 
\draw[color = black, very thick] (-2.1,-2.3) -- (-2.1,-2.6);
\draw[color = black, very thick] (1.7,-2.3) -- (1.4,-2.3);
\draw[color = black, very thick] (1.4,-2.3) -- (1.4,-2.6);
\draw[color = black, very thick] (-4.1,0) -- (-4.1,-.3); 
\draw[color = black, very thick] (-5.5,0) -- (-5.5,-.3); 
\draw[color = black, very thick] (-4.1,-.3) -- (-4.4,-.3) node[left]{};
\draw[color = black, very thick] (-4.4,-.3)--(-4.4,-2.3);
\draw[color = black, very thick] (-5.2,-.3)--(-5.2,-2.3);
\draw[color = black, very thick] (-5.5,-.3) -- (-5.2,-.3) node[left]{};
\draw[color = black, very thick] (-4.1,-2.3) -- (-4.4,-2.3) node[left]{};
\draw[color = black, very thick] (-4.1,-2.3) -- (-4.1,-2.6); 
\draw[color = black, very thick] (-2.1,-2.6) -- (-4.1,-2.6);
\draw[color = black, very thick] (-5.5,-2.3) -- (-5.2,-2.3) node[left]{};
\draw[color = black, very thick] (-5.5,-2.3) -- (-5.5,-2.6); 
\draw[color = black, very thick] (-5.5,-2.6) -- (-6.5,-2.6);
 \end{tikzpicture}   
\caption{Schematic of a lattice of four-waveguide junctions. This problem, which is a possible extension of our work, is motivated by quantum graph theory, as discussed in the introduction. We will calculate the solution in the region in blue. }
\label{graph}
\end{figure}

In this paper, we consider the problem of the scattering of waves by a rectangular region at the junction of four waveguides, which has not been considered previously in the literature. The reason for exploring this problem is partly influenced and motivated by quantum graph theory, in which idealised problems of wave scattering on a metric graph can be formulated and intricate scattering networks can be designed, often with nontrivial properties due to local resonance \cite{Lawrie2023, Lawrie2022}. In this setting, the edges of the graph act as infinitely thin waveguides which support only one mode of propagation. Here, we consider the more general case where the waveguides have a finite thickness and, therefore, support multiple propagating modes. The waveguide junction presented in this paper could be incorporated into a lattice of such junctions, as illustrated in Fig.~\ref{graph}, which could be studied for its locally resonant band structure.

The outline of this paper is as follows: In \textsection \ref{syy}, we formulate the underlying problem, and reduce it to four simpler sub-problems by leveraging the two-fold reflective symmetry of the geometry. In \textsection\ref{KK}, we solve the sub-problems using the eigenfunction matching method (EMM) and present some numerical solutions. In \textsection\ref{scattering}, we describe how our solution can be used to obtain the scattering matrix for the junction, which could be used to solve multiple scattering problems. In \textsection ~\ref{time}, we describe how to compute the time-domain solution numerically from the frequency-domain solutions. Figures showing the time-domain solution for a range of incident pulses and waveguide geometries are given, with corresponding animations being given in the supplementary material. A brief conclusion is given in \textsection \ref{conclusions}. For brevity, some peripheral calculations are relegated to the appendices.

\section{Mathematical Formulation}
\label{syy}
\begin{figure}
   \begin{center}
 \begin{tikzpicture}[x=1.4cm,y=1.4cm][my triangle/.style={-{Triangle[width=\the\dimexpr1.8\pgflinewidth,length=\the\dimexpr0.8\pgflinewidth]}}]
 \fill[cyan!30!white] (0,-0.1) rectangle (-3.5,.3);
  \fill[cyan!30!white] (0,0) rectangle (-1.2,.8);
    \fill[cyan!30!white] (0,0) rectangle (-.7,3);
 \fill[blue!30!white] (3.5,-.5) rectangle (-3.5,-0.1);
  \fill[blue!30!white] (1.2,-1) rectangle (-1.2,-0.1);
  \fill[blue!30!white] (.7,-3) rectangle (-.7,-0.1);
 \fill[blue!30!white] (3.5,-.2) rectangle (0,.3);
\fill[blue!30!white] (1.2,0) rectangle (0,.8);
  \fill[blue!30!white] (.7,0) rectangle (0,3);    
 \draw[->, dashed] (0,-0.1) -- (4,-0.1) node [below] {$x$};    
  \draw[->, dashed] (0,-0.1) -- (0,3.9) node [left] {$y$};    
\usepgflibrary{arrows} 
\draw [teal, -triangle 60 ] (.4,2.6) -- (.4,3.4) node [right] {}; 
\draw [teal, -triangle 60] (.4,-2.6) -- (.4,-3.4) node [right] {}; 
\draw [black, <->, >=triangle 60] (-.5,2.3) -- (.5,2.3) node [right] {};
\node[fill=none, text=black, font=\bfseries, scale=1]at(0,2.5){$2a_2$};
\draw [black, <->, >=triangle 60] (-.5,-2.3) -- (.5,-2.3) node [right] {};
\node[fill=none, text=black, font=\bfseries, scale=1]at(0,-2.1){$2a_2$};
\draw [black, <->, >=triangle 60] (-2.8,-.4) -- (-2.8,.2) node [right] {};
\node[fill=none, text=black, font=\bfseries, scale=1]at(-2.6,-.1){{$2a_1$}};
\draw [black, <->, >=triangle 60] (2.8,-.4) -- (2.8,.2) node [right] {};
\node[fill=none, text=black, font=\bfseries, scale=1]at(2.6,-.1){{$2a_1$}};
\draw [black, <->, >=triangle 60] (-1,-.8) -- (-1,.6) node [right] {};
\node[fill=none, text=black, font=\bfseries, scale=.9]at(-.8,-0.1){{$2b_1$}};
\draw [black, <->, >=triangle 60] (-.8,.6) -- (.8,.6) node [right] {};
\node[fill=none, text=black, font=\bfseries, scale=.9]at(0,0.5){$2b_2$};
\draw [teal, -triangle 60]  (3,-.4) -- (3.8,-.4) node [right] {}; 
\draw [green, -triangle 60]   (-3.8,0)-- (-3,0) node [right] {};
\draw [olive, -triangle 60]  (-3,-.3)--(-3.8,-.3) node [right] {};
\node[color =black, very thick]at(-3.4,.1){$\mathcal{I}$};
\node[color =black, very thick]at(-3.4,-.2){$\mathcal{R}$};
\node[color =black, very thick]at(3.4,-.3){$\mathcal{T}^{\rightarrow}$};
\node[color =black, very thick]at(.2,3.1){$\mathcal{T}^{\uparrow}$};
  \draw[color = black, very thick] (.7,.8)--(.7,3);
\node[color =black, very thick]at(.2,-2.9){$\mathcal{T}^{\downarrow}$};
\draw[color = black, very thick] (1.2,.8) -- (.7,.8) node[left]{};
\draw[color = black, very thick] (1.2,-1) -- (.7,-1) node[left]{};
\draw[color = black, very thick] (-1.2,-1) -- (-.7,-1) node[left]{};
\draw[color = black, very thick] (-.7,.8) -- (-.7,3) node[left]{};
\draw[color = black, very thick] (-.7,-1) -- (-.7,-3) node[left]{};
\draw[color = black, very thick] (.7,-1)--(.7,-3);
\draw[color = black, very thick] (-1.2,.8) -- (-.7,.8); 
\draw[color = black, very thick] (-1.2,.8) -- (-1.2,.3);       
\draw[color = black, very thick] (-1.2,-.5) -- (-3.5,-.5); 
\draw[color = black, very thick] (-1.2,.3) -- (-3.5,.3); 
\draw[color = black, very thick] (1.2,-.5) -- (3.5,-.5); 
\draw[color = black, very thick] (1.2,.3) -- (3.5,.3); 
 \draw[color = black, very thick] (1.2,.8) -- (1.2,.3); 
 \draw[color = black, very thick] (1.2,-.5)--(1.2,-1);
 \draw[color = black, very thick] (-1.2,-.5)--(-1.2,-1);
 \end{tikzpicture}   
   \end{center}
\caption{The waveguide geometry illustrates the incident wave, denoted as $\mathcal{I}$, and the reflected wave, denoted as $\mathcal{R}$. The transmitted waves are represented as $\mathcal{T}^{\uparrow}$, $\mathcal{T}^{\rightarrow}$, and $\mathcal{T}^{\downarrow}$. We have highlighted the principal quadrant in cyan. }
    \label{fig:four}     
\end{figure}
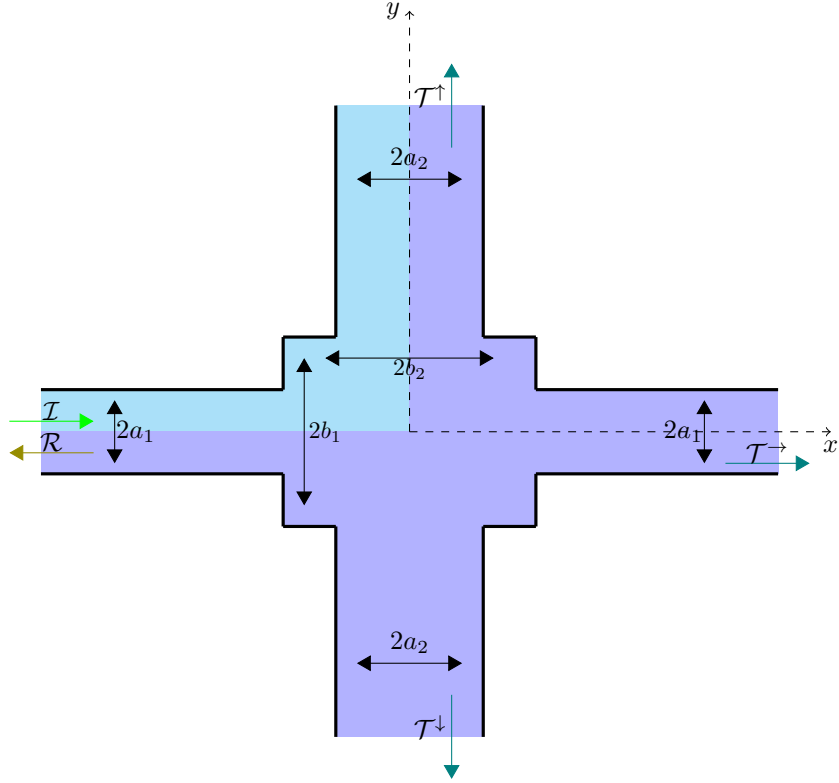

\noindent We consider the non-dimensionalised two-dimensional wave equation given by
\begin{equation}
    \nabla^{2}\varphi(x,y,t) -\frac{\partial^{2}\varphi(x,y,t)}{\partial t^{2}} =0,
    \label{wave equation}
\end{equation}
for $(x,y)\in\Omega$ and $t\in\mathbb{R}$, where $\nabla^{2}$ denotes the Laplacian and $\Omega\subset \mathbb{R}^2$ is the spatial domain. The problem we consider is a two-dimensional waveguide consisting of four channels connected through a rectangular region, as illustrated in Fig.~\ref{fig:four}. The horizontal waveguides each have constant width $2a_1$, and the vertical waveguides each have constant width $2a_2$ and are connected by the rectangular region with height and width $2b_1$ and $2b_2$, respectively. The spatial domain is the following union:

\begin{align}
   \Omega&= \{ (x,y) \mid y \in (-a_1,a_1), x \in (-b_2,-\infty) \}\nonumber \\
   &\quad\cup \{ (x,y) \mid y \in (-b_1,-\infty), x \in (-a_2,a_2) \} \nonumber\\
   &\quad\cup \{ (x,y) \mid y \in (-a_1,a_1), x \in (b_2,\infty) \}  \nonumber  \\
    &\quad\cup \{ (x,y) \mid y \in (b_1,\infty), x \in (-a_2,a_2) \} \nonumber   \\
      &\quad\cup \{ (x,y) \mid y \in (-b_1,b_1), x \in (-b_2,b_2) \}  
\end{align}

Homogeneous Neumann boundary conditions, which describe no-flow boundaries in the context of water waves or sound hard boundaries in the context of acoustics, are prescribed on the boundaries of the waveguide as follows:
\begin{subequations}
\label{33}
\begin{eqnarray}
 \partial_x \varphi&=& 0 , \quad x\in\{-b_2,b_2\}, \quad y \in (-b_1,-a_1) \cup (a_1,b_1),     \\
 \partial_y \varphi&=& 0 , \quad y\in\{-b_1,b_1\}, \quad x \in (-b_2,-a_2) \cup (a_2,b_2),     \\
 \partial_y \varphi&=& 0 , \quad  y\in\{-a_1,a_1\} , \quad x \in [-b_2,-\infty) \cup [b_2,\infty),     \\ 
 \partial_x \varphi&=& 0 , \quad  x\in\{-a_2,a_2\} , \quad y \in [-b_1,-\infty) \cup [b_1,\infty).   
\end{eqnarray}
\end{subequations} 

We express the solution of the wave equation as a continuous superposition of time harmonic functions. The incident wave which is considered here is travelling from left to right
\begin{equation}
  \varphi(x,y,t)= \mathrm{Re} \left \{ \int_{0}^{\infty} \hat f (k) \phi (x,y,k) e^{-\upi k t}  dk\right \},
  \label{time_equation}
\end{equation}
where $\phi(\cdot,\cdot,k)$ satisfies the Helmholtz equation
\begin{equation}\nabla^{2}\phi(x,y)+k^{2}\phi(x,y)= 0,
 \label{helmholtz}
\end{equation}
and inherits the boundary conditions \eqref{wave equation} satisfied by $\varphi$ for all $k>0$. Since wave speed is unity in the non-dimensionalised wave equation, $k$ can be interpreted as both the wavenumber and the angular frequency.

The waveguide has reflective symmetry across the lines $x=0$ and $y=0$, which motivates the decomposition of the problem into four different quadrants, as illustrated in Fig.~\ref{fig:four}. As we will see, the decomposition of the quadrant into rectangular subregions enables a solution to the Helmholtz equation by the method of separation of variables and the eigenfunction matching method.

We now describe how symmetry is used to decompose the full spatial domain into quadrants. If a mode is symmetric about the line $x=0$, then
\begin{equation}
    \phi(x,y)= \phi(-x,y), \quad \forall  (x,y) \in \Omega,
\end{equation}
and its derivative with respect to $x$ must be antisymmetric in $x$, i.e. 
\begin{equation}
    \partial_{x} \phi(x,y)= -   \partial_{x} \phi(-x,y), \quad \forall (x,y) \in \Omega.
\end{equation}
For $x=0$, it follows that
\begin{equation}
     \partial_{x}\phi(0,y)=0, 
\end{equation}
which is a homogeneous Neumann boundary condition. Likewise, if a mode is antisymmetric about the line $x=0$, then
\begin{equation}
    \phi(x,y)=-\phi(-x,y), \quad \forall (x,y) \in \Omega.
\end{equation}
For $x=0$, this implies that
\begin{equation}
 \phi(0,y)=0,   
\end{equation}
which is a homogeneous Dirichlet boundary condition. Similar observations are made about modes that are symmetric/antisymmetric about the line $y=0$. By exploiting symmetry in this way, we will reduce the problem on the full spatial domain to four simpler problems on the principal quadrant $\Omega_P=\{(x,y)\in\Omega|x<0<y\}$. There are:
\begin{itemize}[label=\arabic*)]
    \item [1)] The Neumann-Neumann problem which satisfies homogeneous Neumann boundary conditions at both $x=0$ and $y=0$. We denote the solutions of this problem as $\phi^{\mathcal{NN}}$. See Fig.~\ref{fig:symmetric part}. 
    
    \item [2)] The Dirichlet-Dirichlet problem which satisfies homogeneous Dirichlet boundary conditions at $x=0$ and $y=0$. We denote the solution of this problem as $\phi^{\mathcal{DD}}$. See Fig.~\ref{fig:antisymmetric part}. 
    
    \item [3)] The Neumann-Dirichlet problem which satisfies a homogeneous Dirichlet boundary condition at $x=0$ and a homogeneous Neumann boundary condition at $y=0$. We donate the solution of this problem as $\phi^{\mathcal{ND}}$. See Fig.~\ref{fig:mix part}.  
    
    \item [4)] The Dirichlet-Neumann problem which satisfies a homogeneous Dirichlet boundary condition at $y=0$ and a homogeneous Neumann boundary condition at $x=0$. We donate the solution of this problem as $\phi^{\mathcal{DN}}$. See Fig.~\ref{fig:mix2 part}. 
\end{itemize}

All four problems also satisfy homogeneous Neumann boundary conditions on the physical boundary inherited from \eqref{33}.

We will solve each of the four problems above, assuming a unit-amplitude plane wave incident from $x=-\infty$. The solution for a unit-amplitude plane wave incident from $x=-\infty$ in the full domain can be reconstructed from these solutions in the principal quadrant. If the incident wave is symmetric about the line $y=0$, then the reconstructed solution is
\begin{eqnarray}
\label{11}
 \phi(x,y)= \left\{
\begin{array}{ll}  

\displaystyle\frac{1}{2}(\phi^{\mathcal{NN}}(x,y)+\phi^{\mathcal{D}\mathcal{N}}(x,y)), & \textrm{if} \ x<0,\quad y>0,\\
\displaystyle\frac{1}{2}(\phi^{\mathcal{NN}}(-x,y)-\phi^{\mathcal{D}\mathcal{N}}(-x,y)), & \textrm{if} \  x>0,\quad y>0,\\
\displaystyle\frac{1}{2}(\phi^{\mathcal{NN}}(x,-y)+\phi^{\mathcal{D} \mathcal{N}}(x,-y)), & \textrm{if} \ x<0,\quad y<0,\\
\displaystyle\frac{1}{2}(\phi^{\mathcal{NN}}(-x,-y)-\phi^{\mathcal{D} \mathcal{N}}(-x,-y)), & \textrm{if} \  x>0,\quad y<0.
\end{array} 
\right.    
\label{nn}
\end{eqnarray}
Otherwise, if the incident wave is antisymmetric about the line $x=0$, then the reconstructed solution is
\begin{eqnarray}
\label{22}
 \phi(x,y)= \left\{
\begin{array}{ll}  
\displaystyle\frac{1}{2}(\phi^{\mathcal{DD}}(x,y)+\phi^{\mathcal{N}\mathcal{D}}(x,y)), & \textrm{if} \ x<0,\quad y>0,\\
\displaystyle\frac{1}{2}(\phi^{\mathcal{DD}}(-x,y)-\phi^{\mathcal{N} \mathcal{D}}(-x,y)), & \textrm{if} \  x>0,\quad y>0,\\
\displaystyle\frac{1}{2}(\phi^{\mathcal{DD}}(x,-y)+\phi^{\mathcal{N} \mathcal{D}}(x,-y)), & \textrm{if} \ x<0,\quad y<0,\\
\displaystyle\frac{1}{2}(\phi^{\mathcal{DD}}(-x,-y)-\phi^{\mathcal{N} \mathcal{D}}(-x,-y)), & \textrm{if} \  x>0,\quad y<0.
\end{array} 
\right.    
\label{dd}
\end{eqnarray}

\subsection{Separation of variables}
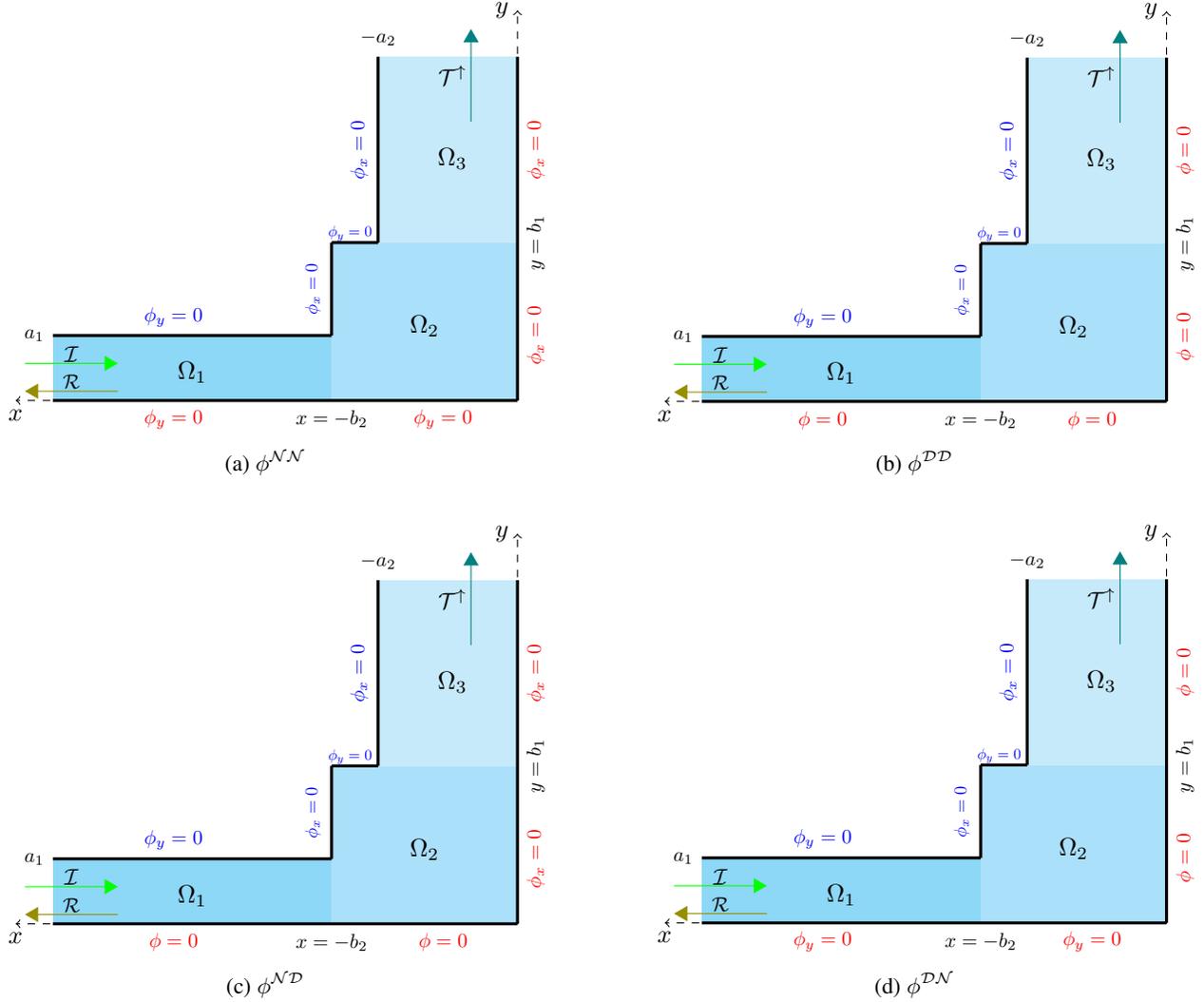
\begin{figure}
    \centering
    \begin{subfigure}[b]{0.45\textwidth}
        \centering
        \begin{tikzpicture}[x=1.3cm,y=1.3cm]
            \fill[cyan!40!white] (-2.9,2.5) rectangle (-6,1.8);   
  \fill[cyan!30!white] (-3,3.5) rectangle (-1,1.8);
    \fill[cyan!20!white] (-2.5,5.5) rectangle (-1,3.5);
   \draw[->, dashed] (-1,1.8) -- (-6.4,1.8) node [below]{$x$};  
  \draw[->, dashed] (-1,1.8) -- (-1,6) node [left]{$y$};    
\usepgflibrary{arrows}
 \draw [olive, -triangle 60] (-5.3,1.9)-- (-6.3,1.9) node [right] {}; 
 \node[fill=none, text=black, font=\bfseries, scale=.8]at(-5.8,2){$\mathcal{R}$};

 \node[color =black, very thick]at(-4.5,2.1){$\Omega_1$};
 \node[color =black, very thick]at(-2,2.6){$\Omega_2$};
 \node[color =black, very thick]at(-1.7,4.4){$\Omega_3$};
 
  \draw [teal, -triangle 60]  (-1.5,4.8)--(-1.5,5.8)node [right] {}; 
  \node[fill=none, text=black, font=\bfseries, scale=.9]at(-1.7,5.3){$\mathcal{T}^{\uparrow}$};
 \draw [green, -triangle 60] (-6.3,2.2) --(-5.3,2.2)node [right] {};  
 \node[fill=none, text=black, font=\bfseries, scale=.9]at(-5.8,2.3){$\mathcal{I}$};
\node[fill=none, text=black, font=\bfseries, scale=.8]at(-6.2,2.5){$a_{1}$};
\node[fill=none, text=black, font=\bfseries, scale=.8]at(-2.5,5.7){$-a_{2}$};
\node[fill=none, text=black, font=\bfseries, scale=.8]at(-.8,3.5){\rotatebox{90}{$y=b_{1}$}};
\node[fill=none, text=red, font=\bfseries, scale=.8]at(-.8,2.5){\rotatebox{90}{$\phi_x=0$}};
\node[fill=none, text=red, font=\bfseries, scale=.8]at(-.8,4.5){\rotatebox{90}{$\phi_x=0$}};
\node[fill=none, text=blue, font=\bfseries, scale=.8]at(-2.7,4.5){\rotatebox{90}{$\phi_x=0$}};
\node[fill=none, text=blue, font=\bfseries, scale=.7]at(-3.2,3){\rotatebox{90}{$\phi_x=0$}};
\node[fill=none, text=blue, font=\bfseries, scale=.6]at(-2.8,3.6){$\phi_y=0$};

\node[fill=none, text=blue, font=\bfseries, scale=.8]at(-4.7,2.7){$\phi_y=0$};
\node[fill=none, text=black, font=\bfseries, scale=.8]at(-3,1.6){$x=-b_2$};
\node[fill=none, text=red, font=\bfseries, scale=.8]at(-4.7,1.6){$\phi_y=0$};
\node[fill=none, text=red, font=\bfseries, scale=.8]at(-1.8,1.6){$\phi_y=0$};
\draw[color = black, very thick] (-1,1.8)--(-1,5.5);
  \draw[color = black, very thick] (-3,2.5)--(-6,2.5);
\draw [color = black, very thick] (-2.5,3.5)--(-3,3.5);
\draw[color =black, very thick] (-2.5,3.5) -- (-2.5,5.5) node[left]{};
\draw[color = black, very thick] (-3,2.5) -- (-3,3.5);       
\draw [color =black, very thick] (-1,1.8) -- (-6,1.8); 
        \end{tikzpicture}
        \caption{$\phi^{\mathcal{NN}}$}
        \label{fig:symmetric part}
    \end{subfigure}
    \hfill
    \begin{subfigure}[b]{0.45\textwidth}
        \centering
        \begin{tikzpicture}[x=1.3cm,y=1.3cm]
           \fill[cyan!40!white] (-2.9,2.5) rectangle (-6,1.8);   
  \fill[cyan!30!white] (-3,3.5) rectangle (-1,1.8);
    \fill[cyan!20!white] (-2.5,5.5) rectangle (-1,3.5);
   \draw[->, dashed] (-1,1.8) -- (-6.4,1.8) node [below]{$x$};  
  \draw[->, dashed] (-1,1.8) -- (-1,6) node [left]{$y$};    
\usepgflibrary{arrows}
 \draw [olive, -triangle 60] (-5.3,1.9)-- (-6.3,1.9) node [right] {}; 
 \node[fill=none, text=black, font=\bfseries, scale=.8]at(-5.8,2){$\mathcal{R}$};

 \node[color =black, very thick]at(-4.5,2.1){$\Omega_1$};
 \node[color =black, very thick]at(-2,2.6){$\Omega_2$};
 \node[color =black, very thick]at(-1.7,4.4){$\Omega_3$};
 
  \draw [teal, -triangle 60]  (-1.5,4.8)--(-1.5,5.8)node [right] {}; 
  \node[fill=none, text=black, font=\bfseries, scale=.9]at(-1.7,5.3){$\mathcal{T}^{\uparrow}$};
 \draw [green, -triangle 60] (-6.3,2.2) --(-5.3,2.2)node [right] {};  
 \node[fill=none, text=black, font=\bfseries, scale=.9]at(-5.8,2.3){$\mathcal{I}$};
\node[fill=none, text=black, font=\bfseries, scale=.8]at(-6.2,2.5){$a_{1}$};
\node[fill=none, text=black, font=\bfseries, scale=.8]at(-2.5,5.7){$-a_{2}$};
\node[fill=none, text=black, font=\bfseries, scale=.8]at(-.8,3.5){\rotatebox{90}{$y=b_{1}$}};
\node[fill=none, text=red, font=\bfseries, scale=.8]at(-.8,2.5){\rotatebox{90}{$\phi=0$}};
\node[fill=none, text=red, font=\bfseries, scale=.8]at(-.8,4.5){\rotatebox{90}{$\phi=0$}};
\node[fill=none, text=blue, font=\bfseries, scale=.8]at(-2.7,4.5){\rotatebox{90}{$\phi_x=0$}};
\node[fill=none, text=blue, font=\bfseries, scale=.7]at(-3.2,3){\rotatebox{90}{$\phi_x=0$}};
\node[fill=none, text=blue, font=\bfseries, scale=.6]at(-2.8,3.6){$\phi_y=0$};

\node[fill=none, text=blue, font=\bfseries, scale=.8]at(-4.7,2.7){$\phi_y=0$};
\node[fill=none, text=black, font=\bfseries, scale=.8]at(-3,1.6){$x=-b_2$};
\node[fill=none, text=red, font=\bfseries, scale=.8]at(-4.7,1.6){$\phi=0$};
\node[fill=none, text=red, font=\bfseries, scale=.8]at(-1.8,1.6){$\phi=0$};
\draw[color = black, very thick] (-1,1.8)--(-1,5.5);
  \draw[color = black, very thick] (-3,2.5)--(-6,2.5);
\draw [color = black, very thick] (-2.5,3.5)--(-3,3.5);
\draw[color =black, very thick] (-2.5,3.5) -- (-2.5,5.5) node[left]{};
\draw[color = black, very thick] (-3,2.5) -- (-3,3.5);       
\draw [color =black, very thick] (-1,1.8) -- (-6,1.8); 
        \end{tikzpicture}
        \caption{$\phi^{\mathcal{DD}}$}
        \label{fig:antisymmetric part}
    \end{subfigure}
    
    \vspace{0.5cm} 
    
    \begin{subfigure}[b]{0.45\textwidth}
        \centering
        \begin{tikzpicture}[x=1.3cm,y=1.3cm]
          \fill[cyan!40!white] (-2.9,2.5) rectangle (-6,1.8);   
  \fill[cyan!30!white] (-3,3.5) rectangle (-1,1.8);
    \fill[cyan!20!white] (-2.5,5.5) rectangle (-1,3.5);
   \draw[->, dashed] (-1,1.8) -- (-6.4,1.8) node [below]{$x$};  
  \draw[->, dashed] (-1,1.8) -- (-1,6) node [left]{$y$};    
\usepgflibrary{arrows}
 \draw [olive, -triangle 60] (-5.3,1.9)-- (-6.3,1.9) node [right] {}; 
 \node[fill=none, text=black, font=\bfseries, scale=.8]at(-5.8,2){$\mathcal{R}$};
 \node[color =black, very thick]at(-4.5,2.1){$\Omega_1$};
 \node[color =black, very thick]at(-2,2.6){$\Omega_2$};
 \node[color =black, very thick]at(-1.7,4.4){$\Omega_3$};
  \draw [teal, -triangle 60]  (-1.5,4.8)--(-1.5,5.8)node [right] {}; 
  \node[fill=none, text=black, font=\bfseries, scale=.9]at(-1.7,5.3){$\mathcal{T}^{\uparrow}$};
 \draw [green, -triangle 60] (-6.3,2.2) --(-5.3,2.2)node [right] {};  
 \node[fill=none, text=black, font=\bfseries, scale=.9]at(-5.8,2.3){$\mathcal{I}$};
\node[fill=none, text=black, font=\bfseries, scale=.8]at(-6.2,2.5){$a_{1}$};
\node[fill=none, text=black, font=\bfseries, scale=.8]at(-2.5,5.7){$-a_{2}$};
\node[fill=none, text=black, font=\bfseries, scale=.8]at(-.8,3.5){\rotatebox{90}{$y=b_{1}$}};
\node[fill=none, text=red, font=\bfseries, scale=.8]at(-.8,2.5){\rotatebox{90}{$\phi_x=0$}};
\node[fill=none, text=red, font=\bfseries, scale=.8]at(-.8,4.5){\rotatebox{90}{$\phi_x=0$}};
\node[fill=none, text=blue, font=\bfseries, scale=.8]at(-2.7,4.5){\rotatebox{90}{$\phi_x=0$}};
\node[fill=none, text=blue, font=\bfseries, scale=.7]at(-3.2,3){\rotatebox{90}{$\phi_x=0$}};
\node[fill=none, text=blue, font=\bfseries, scale=.6]at(-2.8,3.6){$\phi_y=0$};
\node[fill=none, text=blue, font=\bfseries, scale=.8]at(-4.7,2.7){$\phi_y=0$};
\node[fill=none, text=black, font=\bfseries, scale=.8]at(-3,1.6){$x=-b_2$};
\node[fill=none, text=red, font=\bfseries, scale=.8]at(-4.7,1.6){$\phi=0$};
\node[fill=none, text=red, font=\bfseries, scale=.8]at(-1.8,1.6){$\phi=0$};
\draw[color = black, very thick] (-1,1.8)--(-1,5.5);
  \draw[color = black, very thick] (-3,2.5)--(-6,2.5);
\draw [color = black, very thick] (-2.5,3.5)--(-3,3.5);
\draw[color =black, very thick] (-2.5,3.5) -- (-2.5,5.5) node[left]{};
\draw[color = black, very thick] (-3,2.5) -- (-3,3.5);       
\draw [color =black, very thick] (-1,1.8) -- (-6,1.8); 
        \end{tikzpicture}
        \caption{$\phi^{\mathcal{ND}}$}
        \label{fig:mix part}
    \end{subfigure}
    \hfill
    \begin{subfigure}[b]{0.45\textwidth}
        \centering
        \begin{tikzpicture}[x=1.3cm,y=1.3cm]
          \fill[cyan!40!white] (-2.9,2.5) rectangle (-6,1.8);   
  \fill[cyan!30!white] (-3,3.5) rectangle (-1,1.8);
    \fill[cyan!20!white] (-2.5,5.5) rectangle (-1,3.5);
   \draw[->, dashed] (-1,1.8) -- (-6.4,1.8) node [below]{$x$};  
  \draw[->, dashed] (-1,1.8) -- (-1,6) node [left]{$y$};    
\usepgflibrary{arrows}
 \draw [olive, -triangle 60] (-5.3,1.9)-- (-6.3,1.9) node [right] {}; 
 \node[fill=none, text=black, font=\bfseries, scale=.8]at(-5.8,2){$\mathcal{R}$};

 \node[color =black, very thick]at(-4.5,2.1){$\Omega_1$};
 \node[color =black, very thick]at(-2,2.6){$\Omega_2$};
 \node[color =black, very thick]at(-1.7,4.4){$\Omega_3$};
 
  \draw [teal, -triangle 60]  (-1.5,4.8)--(-1.5,5.8)node [right] {}; 
  \node[fill=none, text=black, font=\bfseries, scale=.9]at(-1.7,5.3){$\mathcal{T}^{\uparrow}$};
 \draw [green, -triangle 60] (-6.3,2.2) --(-5.3,2.2)node [right] {};  
 \node[fill=none, text=black, font=\bfseries, scale=.9]at(-5.8,2.3){$\mathcal{I}$};
\node[fill=none, text=black, font=\bfseries, scale=.8]at(-6.2,2.5){$a_{1}$};
\node[fill=none, text=black, font=\bfseries, scale=.8]at(-2.5,5.7){$-a_{2}$};
\node[fill=none, text=black, font=\bfseries, scale=.8]at(-.8,3.5){\rotatebox{90}{$y=b_{1}$}};
\node[fill=none, text=red, font=\bfseries, scale=.8]at(-.8,2.5){\rotatebox{90}{$\phi=0$}};
\node[fill=none, text=red, font=\bfseries, scale=.8]at(-.8,4.5){\rotatebox{90}{$\phi=0$}};
\node[fill=none, text=blue, font=\bfseries, scale=.8]at(-2.7,4.5){\rotatebox{90}{$\phi_x=0$}};
\node[fill=none, text=blue, font=\bfseries, scale=.7]at(-3.2,3){\rotatebox{90}{$\phi_x=0$}};
\node[fill=none, text=blue, font=\bfseries, scale=.6]at(-2.8,3.6){$\phi_y=0$};

\node[fill=none, text=blue, font=\bfseries, scale=.8]at(-4.7,2.7){$\phi_y=0$};
\node[fill=none, text=black, font=\bfseries, scale=.8]at(-3,1.6){$x=-b_2$};
\node[fill=none, text=red, font=\bfseries, scale=.8]at(-4.7,1.6){$\phi_y=0$};
\node[fill=none, text=red, font=\bfseries, scale=.8]at(-1.8,1.6){$\phi_y=0$};
\draw[color = black, very thick] (-1,1.8)--(-1,5.5);
  \draw[color = black, very thick] (-3,2.5)--(-6,2.5);
\draw [color = black, very thick] (-2.5,3.5)--(-3,3.5);
\draw[color =black, very thick] (-2.5,3.5) -- (-2.5,5.5) node[left]{};
\draw[color = black, very thick] (-3,2.5) -- (-3,3.5);       
\draw [color =black, very thick] (-1,1.8) -- (-6,1.8); 
        \end{tikzpicture}
        \caption{$\phi^{\mathcal{DN}}$}
        \label{fig:mix2 part}
    \end{subfigure}
    \caption{Waveguide geometries showing the boundary conditions satisfied by (a) $\phi^{\mathcal{NN}}$, (b) $\phi^{\mathcal{DD}}$, (c) $\phi^{\mathcal{ND}}$, and (d) $\phi^{\mathcal{DN}}$. Physical boundary conditions are donated in blue, whereas artificial boundary conditions are donated in red.}    
    \label{fig:waveguide_subfigures}
\end{figure}

In order to apply the method of separation of variables, the principal quadrant $\Omega_P$ is further decomposed into $\Omega_{1}:= \{ -\infty<x\le-b_2, \quad 0< y \le -a_1\} $,  $\Omega_{2}:= \{0> x \ge -b_2, \quad 0< y \le b_1\} $, and $\Omega_{3}:= \{0> x \ge -a_2, \quad b_1\le y < \infty \} $, as illustrated in Fig.~\ref{fig:symmetric part}. A scattering problem in the quadrant is posed by considering an incident wave propagating along $\Omega_{1}$ from $x=-\infty$ as shown in Fig.~\ref{fig:symmetric part}. The next step is to define the expansions of the eigenfunction in each mode and the orthogonality relation in each region of the channel based on the boundary conditions. Subsequently, these expansions are matched using the following continuity of the pressure and normal velocity conditions:
\begin{subequations}
\begin{eqnarray}
\label{matching1}
\lim_{x\to-b_2^+} \phi^{\mathcal{NN}}(x,y)&=& \lim_{x\to-b_2^-} \phi^{\mathcal{NN}}(x,y), \quad x=-b_2, \quad y \in (0,a_1),\\
    \label{matching2}
    \lim_{y\to b_1^+} \phi^{\mathcal{NN}}(x,y)&=& \lim_{y\to b_1^-} \phi^{\mathcal{NN}}(x,y),  \quad y=b_1, \quad x \in(- a_2, 0). 
\end{eqnarray}
\end{subequations}
and 
\begin{subequations}
\begin{eqnarray}
 \lim_{x\to-b_2^+} \frac{\partial\phi^{\mathcal{NN}}(x,y)}{\partial x} &=& \left\{
\begin{array}{ll}
\label{dreivativex}
   0,  & \textrm{if} \  y \in (a_1, b_1),\\
 \lim_{x\to-b_2^-} \displaystyle \frac{\partial\phi^{\mathcal{NN}}(x,y)}{\partial x},& \textrm{if} \  y \in (0,a_1).
   \end{array} 
\right.   \\  
 \lim_{y\to b_1^+} \frac{\partial\phi^{\mathcal{NN}} (x,y)}{\partial y} &=& \left\{
\begin{array}{ll}
\label{dreivativey}
   0 , & \textrm{if} \   x \in (-b_2,-a_2),\\
\lim_{y\to b_1^-}  \displaystyle \frac{\partial\phi^{\mathcal{NN}}(x,y)}{\partial y}, & \textrm{if} \ x \in ( -a_2,0).
   \end{array} 
\right.  
\end{eqnarray}
\end{subequations}
Identical matching condition hold for $\phi^{\mathcal{DD}}$, $\phi^{\mathcal{ND}}$ and  $\phi^{\mathcal{DN}}$.

Next, the details of the Neumann-Newmann problem will be given in full, with details of the remaining three sub-problems being relegated to \ref{matchingnew}.   
 
\subsection{Neumann-Neumann solutions}
\noindent The Helmholtz equation \eqref{helmholtz} is solved subject to both physical boundary conditions 
\begin{subequations}
\begin{eqnarray}
 \label{condition1}
      \phi_{y}^{\mathcal{NN}}(x,y) &=& 0, \quad y= a_1, b_1    , \quad x \in (-a_2, - \infty), \\
\label{condition2}
  \phi^{\mathcal{NN}}_x(x,y)&=&0,\quad  x=-a_2,-b_2,  \quad y \in (a_1, \infty),
\end{eqnarray}
\end{subequations}
and artificial boundary conditions 
\begin{subequations}
\begin{eqnarray}
  \label{condition3}
  \phi^{\mathcal{NN}}_{y}(x,y)&=& 0, \quad y=0,  \quad x \in (0, -\infty),\\ 
 \label{conditio}
\phi^{\mathcal{NN}} _x(x,y)&=&0,\quad  x=0,  \quad y \in (0, \infty), 
\end{eqnarray}
\end{subequations}

\noindent as shown in Fig.~\ref{fig:symmetric part}. The solution is obtained in piecewise form as
\begin{equation}
\phi^{\mathcal{NN}}(x,y)=\begin{cases}
\phi_1^{\mathcal{NN}}(x,y),&(x,y)\in\Omega_1,\\
\phi^{\mathcal{NN}}_2(x,y)+\phi^{\mathcal{NN}}_{{2}^{'}}(x,y),&(x,y)\in\Omega_2,\\
\phi_3^{\mathcal{NN}},&(x,y)\in\Omega_3.\\
\end{cases}
\end{equation}
The solution in region $\Omega_1$ that satisfies \eqref{helmholtz}, \eqref{condition1} and \eqref{condition2} is written as 
\begin{equation}
  \phi^{\mathcal{NN}}_1(x,y)=e^{i\bar{\beta_{p}} (x+b_{2})} \Phi_p(y)+\sum_{n=0}^{\infty} A^{\mathcal{NN}}_{n} e^{-i\bar{\beta_{n}} (x+b_{2})} \Phi_n(y).   
  \label{incident}
\end{equation}  
Here, $\bar\beta_{n}=\sqrt{k^{2}-\beta_{n}^{2}}$ is the wavenumber of the $n$th reflected mode in which $\beta_{n}=n \pi/a_1$ is the eigenvalue for all $n\in\mathbb{N}_0$. 
The corresponding eigenfunctions $\Phi_n(y)=\cos(\beta_{n}y)$ satisfies the orthogonality relation
\begin{equation}
I_{mn}(a_1)=\int_{0}^{a_1} \Phi_n(y) \Phi_m(y) dy=\kappa_m a_1 \delta_{mn},
\label{U}
   \end{equation}
where $\delta_{mn}$ is Kronecker delta and $\kappa_m$ is defined as 
\begin{eqnarray*}  
 \kappa_m&=& \left\{
\begin{array}{ll}
     1 , & \textrm{if} \ m=n=0,\\
     \frac{1}{2}, & \textrm{if} \ m = n \neq 0,\\
         0 , & \textrm{otherwise}.
   \end{array} 
\right. \\ 
\end{eqnarray*}
Note that the first term in \eqref{incident} describes the incident wave, while the second team describes the reflected wave in which $A^{\mathcal{NN}}_n$ is the amplitude of the $n$th reflected mode. 

In region $\Omega_2$, a superposition of two series is required to satisfy the boundary and matching conditions. Thus, the solutions in region $\Omega_2$ which satisfies \eqref{helmholtz}, \eqref{condition1}, \eqref{condition2} and \eqref{conditio} and then \eqref{helmholtz}, \eqref{condition1}, \eqref{condition2} and \eqref{condition3} respectively are 
\begin{equation}
  \phi^{\mathcal{NN}}_2(x,y)= \sum_{n=0}^{\infty} B^{\mathcal{NN}}_{n} \left( \frac{\cosh(i\bar \alpha_{n}x)}{\cosh(-i\bar \alpha_{n}b_2)}\right) \Gamma_n(y),
  \label{soultion21}
\end{equation}
\begin{equation}
  \phi^{\mathcal{NN}}_{{2}^{'}}(x,y)= \sum_{n=0}^{\infty} C^{\mathcal{NN}}_{n} \left(\frac{\cosh(i\bar  \eta^{'}_n y)}{\cosh(i\bar  \eta^{'}_n b_1)} \right) \Upsilon_n(x).
    \label{soultion22}
\end{equation}
Here, $\bar\alpha_{n}=\sqrt{k^{2}-\alpha_{n}^{2}}$ and $ \bar\eta^{\prime}_n=\sqrt{k^{2}-\eta^{\prime 2}_n}$ are the wavenumbers of the $n$th transmitted mode, in which $\alpha_{n}=n\pi/b_1$ and $\eta^{\prime}_n=n\pi/b_2$ are the eigenvalues for $ n=0,1,2,\dots$. The corresponding eigenfunctions $\Gamma_n(y)=\cos (\alpha_{n}y)$ and $\Upsilon_n(x)=\cos(\eta^{'}_n x)$ satisfy the orthogonality relation on their respective domains of definition.  The solution in region $\Omega_3$ that satisfies \eqref{helmholtz}, \eqref{condition3}, and \eqref{conditio} is      
\begin{equation}
  \phi^{\mathcal{NN}}_3(x,y)= \sum_{n=0}^{\infty} D^{\mathcal{NN}}_n e^{i\bar{\eta_{n}}(y-b_1)}\aleph_{n}(x). 
  \label{solution31}
\end{equation}
Here $\bar\eta_{n}=\sqrt{k^{2}-\eta_{n}^{2}}$ is the wavenumber of the $n$th transmitted mode in which $\eta_{n}= n\pi/a_2$ is the eigenvalue for $n\in\mathbb{N}_0$. The corresponding eigenfunction $\aleph_{n}(x)=\cos \eta_{n} (x)$ satisfies the orthogonality on their respective domains of definition.
Note that $D^{\mathcal{NN} }_n$ in \eqref{solution31} is the amplitude of the transmitted mode coefficient.

\subsection{Dirichlet-Dirichlet solutions}
\noindent The Helmholtz equation \eqref{helmholtz} is solved subject to physical 
boundary conditions 
\begin{subequations}
\begin{eqnarray}
\label{condition4}
 \phi^{\mathcal{DD}}_{y}(x,y)&=& 0, \quad y= a_1,b_1 , \quad  x \in ( -a_2, - \infty),\\
  \label{condition5}    
 \phi^{\mathcal{DD}}(x,y)&=& 0, \quad y= 0,  \quad  x \in (0, -\infty),
\end{eqnarray}
\end{subequations}
and artificial boundary conditions   
\begin{subequations}
\begin{eqnarray} 
 \label{condition6}
 \phi^{\mathcal{DD}}_x(x,y)&=& 0,\quad  x=-b_2,-a_2, \quad  y \in ( a_1, \infty)\\  
 \label{conditio1}
\phi^{\mathcal{DD}}(x,y)&=& 0,\quad  x=0,  \quad y\in ( 0, \infty),  
\end{eqnarray}
\end{subequations}

\noindent as shown in Fig.~\ref{fig:antisymmetric part}. The solution, in this case, is obtained in piecewise form as
\begin{equation}
\phi^{\mathcal{DD}}(x,y)=\begin{cases}
\phi_1^{\mathcal{DD}}(x,y),&(x,y)\in\Omega_1,\\
\phi^{\mathcal{DD}}_2(x,y)+\phi^{\mathcal{DD}}_{{2}^{'}}(x,y),&(x,y)\in\Omega_2,\\
\phi_3^{\mathcal{DD}},&(x,y)\in\Omega_3.\\
\end{cases}
\end{equation}

\noindent The solution in region $\Omega_1$ which satisfies \eqref{helmholtz}, \eqref{condition4}, and \eqref{condition5} is 
\begin{equation}
\phi^{\mathcal{DD}}_1(x,y)=e^{i\bar{\gamma_{p}} (x+b_{2})}\widetilde \Phi_p(y) +\sum_{n=1}^{\infty} A^{\mathcal{DD}}_{n} e^{-i\bar{\gamma_{n}} (x+b_{2})} \widetilde \Phi_n(y).
\label{incident2}
\end{equation}  
Here, $\bar\gamma_{n}=\sqrt{k^{2}-\gamma_{n}^{2}}$, $\gamma_{n}=(2n+1)\pi/2a_1 $ for $n=0,1,2,\dots$ and $\widetilde \Phi_n(y)=\sin(\gamma_{n}y)$ satisfy the orthogonality  
\begin{eqnarray}
        \hat I_{mn}(a_1)=\int_{0}^{a_1} \widetilde \Phi_n(y) \widetilde \Phi_m(y) dy=\frac{1}{2} a_1 \delta_{mn},
    \label{hat u}
\end{eqnarray}
The first term in \eqref{incident2} is the incident wave, and the second term is the reflected field in which $A^{\mathcal{DD}}_n$ is the amplitude of the $n$th reflected mode. The solutions in region $\Omega_2$ that satisfy \eqref{helmholtz}, \eqref{condition4}, \eqref{condition5}, and \eqref{conditio1} and then  \eqref{helmholtz}, \eqref{condition5}, \eqref{condition6} and \eqref{conditio1} respectively are
\begin{equation}
  \phi^{\mathcal{DD}}_2(x,y)= \sum_{n=1}^{\infty} B^{\mathcal{DD}}_{n} \left( \frac{\sinh(i\bar \lambda_{n}x)}{\sinh(-i\bar \lambda_{n}b_2)}\right) \widetilde \Gamma_n(y),
  \label{solutionanti1}
\end{equation}
\begin{equation}
  \phi^{\mathcal{DD}}_{{2}^{\prime}}(x,y)= \sum_{n=1}^{\infty} C^{\mathcal{DD}}_{n} 
  \left(\frac{\sinh(i\bar\zeta_{n}y)}{\sinh(i\bar\zeta_{n}b_1)} \right) \widetilde \Upsilon_n(x),
   \label{solutionanti2}
\end{equation}
where $\bar\lambda_{n}=\sqrt{k^{2}-\lambda_{n}^{2}}$, $\bar \zeta_{n}=\sqrt{k^{2}-\zeta_{n}^{2}}$, $\lambda_{n}=(2n+1)\pi/2b_2$, $\zeta_{n}=(2n+1)\pi/2b_2$ and $\widetilde\Gamma_n(y)=\sin (\lambda_{n}y)$ and $\widetilde\Upsilon_n(x)=\sin (\zeta_{n}y)$ satisfy the orthogonality. 
The solution in region $\Omega_3$ that satisfy \eqref{helmholtz}, \eqref{condition6} , and \eqref{conditio1} is 
\begin{equation}
  \phi^{\mathcal{DD}}_3(x,y)= \sum_{n=1}^{\infty} D^{\mathcal{DD}}_n e^{i\bar{\zeta^{\prime}_{n}}(y-b_1)}\widetilde \aleph_{n}(x),
  \label{soltuionant3}
\end{equation}
where $\bar\zeta^{\prime}_{n}=\sqrt{k^{2}-\bar\zeta^{\prime2}_{n}}$, $\bar\zeta^{\prime}_{n}=(2n+1)\pi/2a_2$, for $n=0,1,2,\dots$ and $\widetilde \aleph_{n}(x)=\sin (\zeta^{\prime}_{n}x)$ satisfy the orthogonality.

\noindent Note that $D^{\mathcal{DD}}_n$ in \eqref{soltuionant3} is the amplitude of the transmitted mode coefficient.

\subsection{Neumann-Dirichlet solutions}
\noindent The Helmholtz equation \eqref{helmholtz} is solved subject to physical boundary conditions 
\begin{subequations}
\begin{eqnarray}
\label{condition222}  
\phi_y(x,y)&=& 0,\quad  y=b_1,  \quad  x \in (-b_2,-a_2),\\
\label{condition222_1} 
\phi_x(x,y)&=& 0,\quad  x=-b_2,  \quad y \in (a_1,b_1),
\end{eqnarray}
\end{subequations}
and artificial boundary conditions 
\begin{subequations}
\begin{eqnarray}
\label{condition222_11} 
\phi_x(x,y)&=& 0,\quad  x=0,  \quad y \in (0,b_1),\\
 \label{condition222_2}  
\phi(x,y)&=& 0, \quad y= 0,  \quad x \in (0,-b_2). 
\end{eqnarray}
\end{subequations}

\noindent  as shown in Fig.~\ref{fig:mix part}. The solution in $\Omega_1$ is of the form given in \eqref{incident2}. The solutions in region $\Omega_2$ that satisfy \eqref{helmholtz}, \eqref{condition222}, \eqref{condition222_1}, and \eqref{condition222_2} and then satisfy \eqref{helmholtz}, \eqref{condition222}, \eqref{condition222_1}, and \eqref{condition222_11} are 
\begin{equation}
  \phi^{\mathcal{N}\mathcal{D}}_2(x,y)= \sum_{n=0}^{\infty} B^{\mathcal{N}\mathcal{D}}_{n} \left( \frac{\cosh(i\bar \lambda_{n}x)}{\cosh(-i\bar \lambda_{n}b_2)}\right) \widetilde\Gamma_{n}(y),
  \label{mix1}
\end{equation}
\begin{equation}
  \phi^{\mathcal{N}\mathcal{D}}_{{2}^{\prime}}(x,y)= \sum_{n=1}^{\infty} C^{\mathcal{N}\mathcal{D}}_{n} \left(\frac{\sinh(i\bar  \eta^{\prime}_n y)}{\sinh(i\bar  \eta^{\prime}_n b_1)} \right) \Upsilon_{n}(x),
  \label{mix2}
\end{equation}
Additionally, the solution in $\Omega_3$ is of the form given in \eqref{solution31}, where the definitions of the eigenvalues and the eigenfunctions remain unchanged.

\subsection{Dirichlet-Neumann solutions}
\noindent The Helmholtz equation \eqref{helmholtz} is solved subject to physical 
boundary conditions 
\begin{subequations}
\begin{eqnarray}
\label{condition555} 
\phi_{y}(x,y)&=& 0, \quad y=b_1,  \quad x \in (-b_2,-a_2), \\
\label{condition555_1} 
\phi_x(x,y)&= & 0,\quad  x=-b_2,  \quad  y \in (a_1,b_1),
\end{eqnarray}
\end{subequations}
and artificial boundary conditions   

\begin{subequations}
\begin{eqnarray}
\label{condition555_111} 
\phi_{y}(x,y)&=& 0, \quad y=0,  \quad x \in (-b_2,0), \\
\label{condition555_2} 
\phi(x,y)&=& 0,\quad  x=0,  \quad y \in (b_1,0).
\end{eqnarray}
\end{subequations}

\noindent as shown in Fig.~\ref{fig:mix2 part}. The solution in $\Omega_1$ is of the form given in \eqref{incident}. The solutions in region $\Omega_2$ that satisfy \eqref{helmholtz}, \eqref{condition555}, \eqref{condition555_111}, and ~\eqref{condition555_2} and then satisfy \eqref{helmholtz}, \eqref{condition555}, \eqref{condition555_1}, and ~\eqref{condition555_2} are 
\begin{equation}
  \phi^{\mathcal{D}\mathcal{N}}_2(x,y)= \sum_{n=0}^{\infty} B^{\mathcal{D}\mathcal{N}}_{n} \left( \frac{\sinh(i\bar \alpha_{n}x)}{\sinh(-i\bar \alpha_{n}b_2)}\right) \Gamma_{n}(y),
  \label{mix3}
\end{equation}
\begin{equation}
  \phi^{\mathcal{D}\mathcal{N}}_{{2}^{\prime}}(x,y)= \sum_{n=0}^{\infty} C^{\mathcal{D}\mathcal{N}}_{n} \left(\frac{\cosh(i\bar\zeta_{n}y)}{\cosh(i\bar\zeta_{n}b_1)} \right)\widetilde\Upsilon_{n}(x). 
  \label{mix4}
\end{equation} 
The solution in $\Omega_3$ is of the form given in \eqref{soltuionant3}.
Where there is no change in the eigenvalues and the eigenfunctions. 

Next, we will describe how the unknown coefficients are determined numerically using the eigenfunction matching method. 

\section{Numerical solution using eigenfunction matching}
\label{KK}

\noindent Details of the eigenfunction matching method are given in full for the Neumann-Neumann problem, with those of the remaining three subproblems being relegated to the appendix. After substituting the general solutions \eqref{incident}, \eqref{soultion21} and \eqref{soultion22}, the matching conditions at the interface $x=-b_2$, \eqref{matching1} becomes
\begin{equation*}
\Phi_p(y)+\sum_{n=0}^{\infty} A^{\mathcal{NN}}_{n} \Phi_n(y)=\sum_{n=0}^{\infty} B^{\mathcal{NN}}_{n} \Gamma_n(y),+ \sum_{n=0}^{\infty} C^{\mathcal{NN}}_{n} \left(\frac{\cosh(i\bar\eta^{\prime}_n y)}{\cosh(i\bar  \eta^{\prime}_n b_1)} \right) \cos(-\pi n).     
\end{equation*}
Taking the inner product of the above with $ \Phi_m(y)$ and integrating over $[0,a_1]$, then applying \eqref{U}, yields an expression of the form
\begin{equation}
 I_{mp}(a_1)+A^{\mathcal{NN}}_{m} U_{mn}=\sum_{n=0}^{\infty} B^{\mathcal{NN}}_{n} H_{mn}+\sum_{n=0}^{\infty} C^{\mathcal{NN}}_{n} V_{mn}, \quad m=0,1,2,\dots. 
  \label{system1}
\end{equation}
We defer the statement of the quantities $H_{mn}$ and $V_{mn}$ to Appendices \ref{H} and \ref{V}.

At the second interface $y=b_1$, we substitute \eqref{solution31}, \eqref{soultion21} and \eqref{soultion22} in the matching condition \eqref{matching2} to obtain   
\begin{equation*}
\sum_{n=0}^{\infty} D^{\mathcal{NN}}_n  \aleph_{n}(x)= \sum_{n=0}^{\infty} B^{\mathcal{NN}}_{n} \left( \frac{\cosh(i\bar \alpha_{n}x)}{\cosh(-i\bar \alpha_{n}b_2)}\right) \cos(n\pi)+ \sum_{n=0}^{\infty} C^{\mathcal{NN}}_{n} \Upsilon_n(x).  
\end{equation*}
Taking the inner product with $ \aleph_{m}(x)$ and integrating over $[-a_2,0]$, then applying the orthogonality, yields an expression of the form
\begin{equation}
   D^{\mathcal{NN}}_{m} I_{mn}(a_2)=\sum_{n=0}^{\infty} B^{\mathcal{NN}}_{n} O_{mn}+\sum_{n=0}^{\infty} C^{\mathcal{NN}}_{n} E_{mn}, \quad m=0,1,2,\dots,   
    \label{system2}
\end{equation}
where $O_{mn}$ and $E_{mn}$ are given in Appendices \ref{O} and \ref{E}.

Now, we apply the derivative matching conditions. At $x=-b_2$, use of the derivative matching condition \eqref{dreivativex} yields
\begin{equation}
     \bar\beta_{p} \Phi_p(y)- \sum_{n=0}^{\infty}  \bar\beta_{n} A^{\mathcal{NN}}_{n} \Phi_n(y)= \sum_{n=0}^{\infty} B^{\mathcal{NN}}_{n} \bar\alpha_{n}\tanh(-i\bar\alpha_{n}b_{2}) \Gamma_n(y).
\end{equation}
Taking the inner product with $\Gamma_m(y)$ and integrating over $[0,b_1]$, then using \eqref{H}, yields  
\begin{equation}
\bar \beta_p H_{pm}- \sum_{n=0}^{\infty} A^{\mathcal{NN}}_{n} \bar\beta_{n} H_{nm}=  B^{\mathcal{NN}}_{m} \bar\alpha_{m} \tanh(-i\bar\alpha_{m}b_{2}) I_{mn}(b_1), \quad m=0,1,2,\dots.
\label{system3}
\end{equation}
At the second discontinuity $y=b_1$, \eqref{dreivativey} yields
\begin{equation}
     \sum_{n=0}^{\infty} A^{\mathcal{NN}}_{n} \bar\eta_{n} \aleph_n(x)=\sum_{n=0}^{\infty} B^{\mathcal{NN}}_{n} \bar\eta^{\prime}_{n}  \tanh(i\bar\eta^{\prime}_{n}b_{1})\Upsilon_{n}(x).
\end{equation}
Taking the inner product with $\Upsilon_{m}(x)$ and integrating  over $[-b_2,0]$, then using \eqref{E}, yields
\begin{equation}
    \sum_{n=0}^{\infty} D^{\mathcal{NN}}_{n} \bar\eta_{n}  E_{nm} = C^{\mathcal{NN}}_{m} \bar\eta^{\prime}_{m}  \tanh(i\bar\eta^{\prime}_{m}b_{1}) I_{mn}(b_2), \quad m=0,1,2,\dots. 
    \label{system4}
\end{equation}

Continue with the calculations in the preceding subsection. We truncate the summation, and then we express the four systems of equations in a block matrix \eqref{system1}, \eqref{system2}, \eqref{system3} and ~\eqref{system4} as

{\footnotesize
\begin{equation*}
\begin{array}{c}
    \begin{bmatrix}
        \left[-\textrm{diag}(I_{mn}(a_1)) \right] & \left[H_{mn}\right] & \left[V_{mn}\right] & \left[0\right] \\
        \left[0\right] & \left[O_{mn}\right] & \left[E_{mn}\right] & \left[-\textrm{diag}(I_{mn}(a_2))\right] \\
        \left[\textrm{diag}(\bar\beta_{n}) H^{T}_{mn} \right] & \left[\textrm{diag}(\bar\alpha_{m} \tanh(-i\bar\alpha_{m}b_{2}) 
        I_{mn}(b_1)) \right] & \left[0\right] & \left[0\right] \\
        \left[0\right] & \left[0\right] & \left[\textrm{diag}(\bar\eta^{\prime}_{m} \tanh(i\bar\eta^{\prime}_{m}b_{1}) I_{mn}(b_2))\right] & \left[-\textrm{diag}(\bar\eta_{n}) E^{T}_{mn}\right]
    \end{bmatrix}\\
    \begin{bmatrix}
        \left[A^{\mathcal{NN}}_n\right] \\
        \left[B^{\mathcal{NN}}_n\right] \\
        \left[C^{\mathcal{NN}}_n\right] \\
        \left[D^{\mathcal{NN}}_n\right]
    \end{bmatrix}
    =
    \begin{bmatrix}
        \left[I_{mn}(a_1)\right] \\
        \left[0\right] \\
        \left[\bar\beta_{p}H^{T}_{mp}\right] \\
        \left[0\right]
    \end{bmatrix}.
\end{array}
\end{equation*}}

where the outer bracket denotes the matrix, the above-linear system of equations is solved numerically using MATLAB to define the unknown quantities. Then, the plots for the matching pressure and normal velocity conditions are provided in Supplementary Materials. Here, we present the surface of the solutions.

\begin{figure}
    \centering
    \begin{subfigure}{0.45\textwidth}
        \centering
        \includegraphics[width=\linewidth]{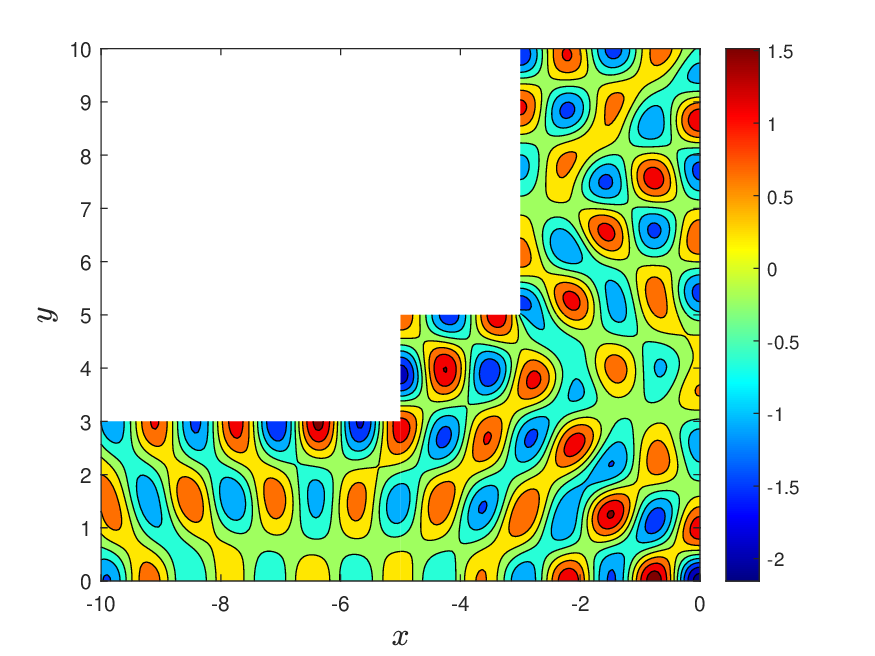}
        \caption{$\phi^{\mathcal{NN}}$}
        \label{symm3}
    \end{subfigure}
    \hfill
    \begin{subfigure}{0.45\textwidth}
        \centering
        \includegraphics[width=\linewidth]{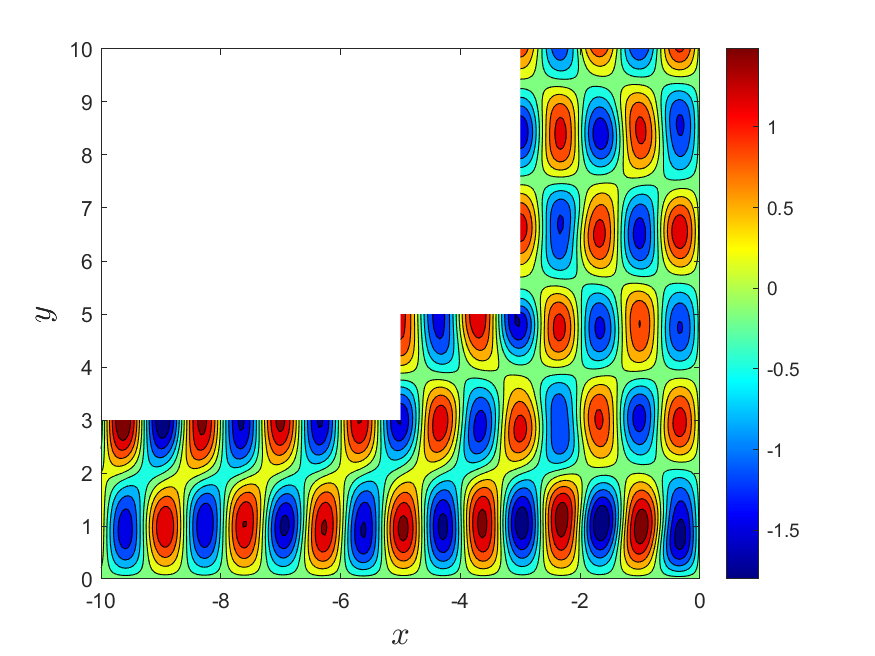}
        \caption{$\phi^{\mathcal{DD}}$}
        \label{anti3}
    \end{subfigure}
    
    \vspace{0.5cm} 

    \begin{subfigure}{0.45\textwidth}
        \centering
        \includegraphics[width=\linewidth]{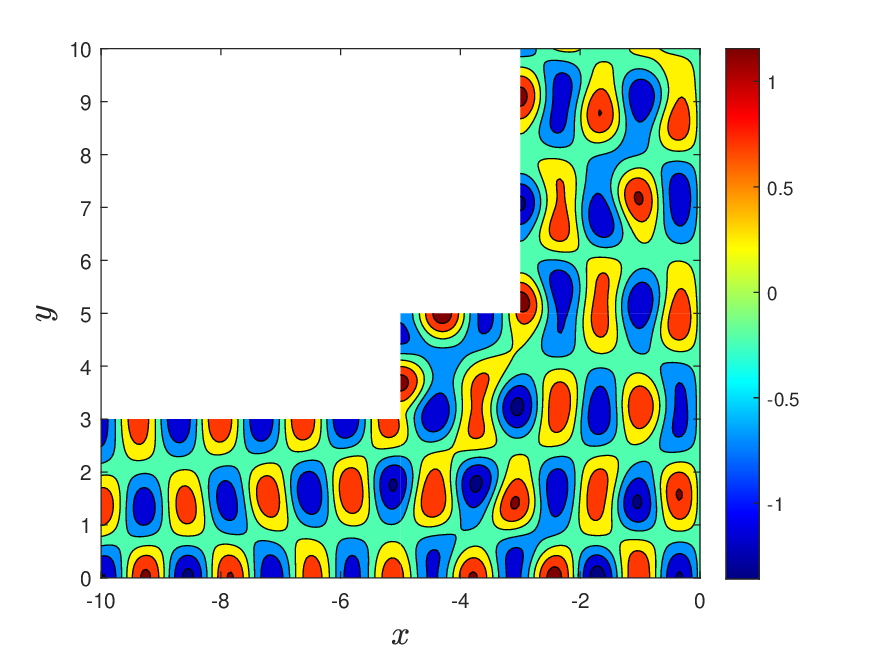}
        \caption{$\phi^{\mathcal{ND}}$}
        \label{mix13}
    \end{subfigure}
    \hfill
    \begin{subfigure}{0.45\textwidth}
        \centering
        \includegraphics[width=\linewidth]{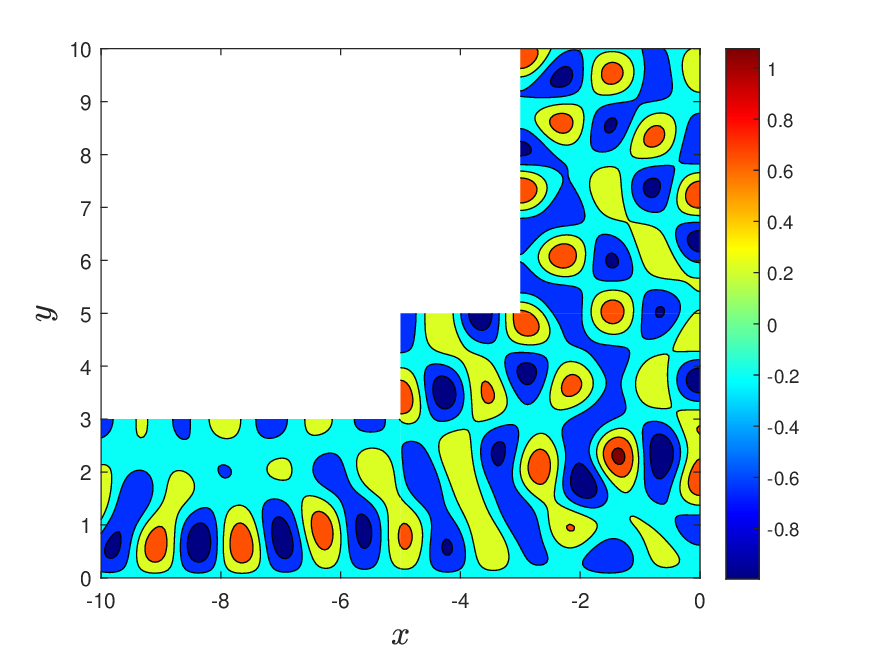}
        \caption{$\phi^{\mathcal{DN}}$}
        \label{mix23}
    \end{subfigure}
    
    \caption{Contour plots of the real parts of (a) $\phi^{\mathcal{NN}}$, (b) $\phi^{\mathcal{DD}}$, 
    (c) $\phi^{\mathcal{ND}}$, and (d) $\phi^{\mathcal{DN}}$ for the cases where $a_1 = a_2 = 3$, 
    $b_1 = b_2 = 5$, $k = 5$, and $N = 100$. The incident mode is $p=2$ in panels (a) and (c), 
    and $p=1$ in panels (b) and (d). Note that contour lines being parallel or perpendicular 
    to a boundary indicate homogeneous Dirichlet or Neumann boundary conditions, respectively.}
    \label{mix23}
\end{figure}

Figure \ref{mix23} shows numerical results for the four quadrant problems for the case $a_{1}=a_{2}=3$, $b_{1}=b_{2}=5$ and $k=5$, with the truncation value $N=100$. This truncation value was chosen so that (i) the matching conditions were satisfied to linewidth, and (ii) the appropriate conservation of energy identity holds numerically. In the Neumann-Neumann case where $a_1 = a_2$, these are given by
\begin{subequations}
\begin{equation}
 (|A^{\mathcal{NN}}_{0}|^{2}+| D^{\mathcal{NN}}_{0}|^{2})+ \displaystyle \frac{1}{2}\sum_{n=1}^{q} \left(\frac{\bar\beta_{n}}{ \bar\beta_{0}} \right)( |A^{\mathcal{NN}}_{n}|^{2}+|D^{\mathcal{NN}}_{n}|^{2})=1,   
\end{equation} 
for $p=0$, and by 
\begin{equation}
  \displaystyle  \left(\frac{2 \bar\beta_{0}}{\bar\beta_{p} } \right) (|A^{\mathcal{NN}}_{0}|^{2}+| D^{\mathcal{NN}}_{0}|^{2})+ \displaystyle \sum_{n=1}^{q} \left(\frac{\bar\beta_{n}}{ \bar\beta_{p}} \right)( |A^{\mathcal{NN}}_{n}|^{2}+|D^{\mathcal{NN}}_{n}|^{2})= 1, 
\end{equation}
\end{subequations}
for $p\neq 0$ where $q = \left\lfloor \frac{ka_1}{\pi} \right\rfloor$ is the number of even propagating modes.

\section{The Scattering Matrix}
\label{scattering}

\noindent Here, we outline a scattering matrix for the four-waveguide problem, which could be used to extend our methods to multiple scattering problems, such as the one illustrated in Figure \ref{graph}. For simplicity, we only consider the case $a_1=a_2$ and $b_1=b_2$, although our discussion can be extended to the more general case. The scattering matrix satisfies
 
\begin{equation}
 \mathcal{S} \mathbf{a} = \mathbf{b},
\end{equation}
where the entries of $\mathbf{a}$ are the coefficients of the incident wave entering the channels, and the entries of $\mathbf{b}$ are the coefficients of the outgoing waves. We express these vectors as  

\[
\mathbf{a}=  \begin{bmatrix}
 \mathbf{a}_{\text{even}, {\rightarrow}}  \\ 
 \mathbf{a}_{\text{odd}, {\rightarrow}} \\
 \mathbf{a}_{\text{even}, {\downarrow}} \\
 \mathbf{a}_{\text{odd}, {\downarrow}} \\ 
 \mathbf{a}_{\text{even}, {\leftarrow}}  \\
 \mathbf{a}_{\text{odd}, {\leftarrow}}  \\
 \mathbf{a}_{\text{even}, {\uparrow}} \\ 
 \mathbf{a}_{\text{odd}, {\uparrow}} \\
\end{bmatrix},
\quad \text{and} \quad
\mathbf{b}=  \begin{bmatrix}
\mathbf{b}^{\text{even},  {\leftarrow}}   \\ 
\mathbf{b} ^{\text{odd}, {\leftarrow}} \\
\mathbf{b} ^{\text{even}, {\uparrow}}   \\
\mathbf{b}^{\text{odd},  {\uparrow}} \\ 
\mathbf{b}^{\text{even},  {\rightarrow}}   \\
\mathbf{b}^{\text{odd},  {\rightarrow}}  \\
\mathbf{b}^{\text{even},  {\downarrow}}  \\ 
\mathbf{b}^{\text{odd}, {\downarrow}} \\
\end{bmatrix}
\]
where $\mathbf{a}_{\text{even}, {\rightarrow}}$, $  \mathbf{a}_{\text{even}, {\downarrow}}$, $ \mathbf{a}_{\text{even}, {\leftarrow}}$ and $ \mathbf{a}_{\text{even}, {\uparrow}}$ denote the amplitude coefficients of even incident waves entering from the left, top, right and bottom channels, respectively, whereas $\mathbf{b}^{\text{even},{\leftarrow}}$, $  \mathbf{b}^{\text{even}, {\uparrow}}$, $ \mathbf{b}^{\text{even}, {\rightarrow}}$ and $ \mathbf{b}^{\text{even}, {\downarrow}}$ denote the amplitude coefficients of even outgoing waves exiting through the left, top, right and bottom channels, respectively. The remaining subvectors of $\mathbf{a}$ and $\mathbf{b}$, which correspond to waves with odd symmetry, are defined similarly. We assume that the wide spacing approximation will be applied to the ensuing multiple scattering problem, so evanescent incident/outgoing wave components are not considered. Thus, the vectors of even wave coefficients are of length $q+1$, where we recall $q = \left\lfloor \frac{ka_1}{\pi} \right\rfloor$, and the vectors of odd wave coefficients are of length $\tilde{q}= \left\lfloor \frac{2ka_1-\pi}{2\pi} \right\rfloor$, being the number of odd propagating modes. The scattering matrix $\mathcal{S}$ is expressed as an $8\times 8$ block matrix of the form                                                 
\begin{equation*}
\mathcal{S}=
\begin{array}{cccc}
\left[ \begin{array}{cccccccccccc}
S^{\text{even},\leftarrow}_{\text{even},\rightarrow}&
[0]&
S^{\text{even},\leftarrow}_{\text{even},\downarrow}&
S^{\text{even},\leftarrow}_{\text{odd},\downarrow}&
S^{\text{even},\leftarrow}_{\text{even},\leftarrow}& 
[0]&
S^{\text{even},\leftarrow}_{\text{even},\uparrow}& S^{\text{even},\leftarrow}_{\text{odd},\uparrow}
\\

[0]&
S^{\text{odd},\leftarrow}_{\text{odd},\rightarrow}&
S^{\text{odd},\leftarrow}_{\text{even},\downarrow}&
S^{\text{odd},\leftarrow}_{\text{odd},\downarrow}&
[0]&
S^{\text{odd},\leftarrow}_{\text{odd},\leftarrow}&
S^{\text{odd},\leftarrow}_{\text{even},\uparrow}&
S^{\text{odd},\leftarrow}_{\text{odd},\uparrow}
\\

S^{\text{even},\uparrow}_{\text{even},\rightarrow}& 
S^{\text{even},\uparrow}_{\text{odd},\rightarrow}&
S^{\text{even},\uparrow}_{\text{even},\downarrow}&
[0]&
S^{\text{even},\uparrow}_{\text{even},\leftarrow}&
S^{\text{even},\uparrow}_{\text{odd},\leftarrow}&
S^{\text{even},\uparrow}_{\text{even},\uparrow}& 
[0]
\\

S^{\text{odd},\uparrow}_{\text{even},\rightarrow}& 
S^{\text{odd},\uparrow}_{\text{odd},\rightarrow}&
[0]&
S^{\text{odd},\uparrow}_{\text{odd},\downarrow}&
S^{\text{odd},\uparrow}_{\text{even},\leftarrow}&
S^{\text{odd},\uparrow}_{\text{odd},\leftarrow}&
[0]& 
S^{\text{odd},\uparrow}_{\text{odd},\uparrow}
\\

S^{\text{even},\rightarrow}_{\text{even},\rightarrow}& 
[0]&
S^{\text{even},\rightarrow}_{\text{even},\downarrow}&
S^{\text{even},\rightarrow}_{\text{odd},\downarrow}&
S^{\text{even},\rightarrow}_{\text{even},\leftarrow}&
[0]&
S^{\text{even},\rightarrow}_{\text{even},\uparrow}& 
S^{\text{even},\rightarrow}_{\text{odd},\uparrow}
\\

[0]& 
S^{\text{odd},\rightarrow}_{\text{odd},\rightarrow}&
S^{\text{odd},\rightarrow}_{\text{even},\downarrow}&
S^{\text{odd},\rightarrow}_{\text{odd},\downarrow}&
[0]&
S^{\text{odd},\rightarrow}_{\text{odd},\leftarrow}&
S^{\text{odd},\rightarrow}_{\text{even},\uparrow}& 
S^{\text{odd},\rightarrow}_{\text{odd},\uparrow}
\\

S^{\text{even},\downarrow}_{\text{even},\rightarrow}& 
S^{\text{even},\downarrow}_{\text{odd},\rightarrow}&
S^{\text{even},\downarrow}_{\text{even},\downarrow}&
[0]&
S^{\text{even},\downarrow}_{\text{even},\leftarrow}&
S^{\text{even},\downarrow}_{\text{odd},\leftarrow}&
S^{\text{even},\downarrow}_{\text{even},\uparrow}& 
[0]
\\

S^{\text{odd},\downarrow}_{\text{even},\rightarrow}& 
S^{\text{odd},\downarrow}_{\text{odd},\rightarrow}&
[0]&
S^{\text{odd},\downarrow}_{\text{odd},\downarrow}&
S^{\text{odd},\downarrow}_{\text{even},\leftarrow}&
S^{\text{odd},\downarrow}_{\text{odd},\leftarrow}&
[0]& 
S^{\text{odd},\downarrow}_{\text{odd},\uparrow}
\\

\end{array} \right].
\end{array}
\end{equation*}
where the notation of the submatrices is defined so that, for example, $S^{\text{odd},\uparrow}_{\text{even},\rightarrow}$ gives the coefficients of odd outgoing waves travelling upwards towards $y=+\infty$ due to excitation by even incident waves travelling right from $x=-\infty$. The scattering matrix contains zero matrices $[0]$ due to symmetry restrictions, as for example, a wave of even symmetry incident from $x=-\infty$ cannot excite waves of odd symmetry outgoing towards $x=\pm\infty$.

In the case where $k=4$ and $a_1=a_2=2$, there are three propagating symmetric modes and two propagating antisymmetric modes in each channel. This means that vectors of the form $\mathbf{a}_{\text{even}}$ and $\mathbf{b}^{\text{even}}$ are of length q+1=3, while vectors of the form $\mathbf{a}_{\text{odd}}$ and $\mathbf{b}^{\text{odd}}$ are of length $\tilde{q}=2$. The entries of the nonzero submatrices of $\mathcal{S}$ can be computed from our solution. For brevity, only two examples are given here. The matrix 
$S^{\text{even},\leftarrow}_{\text{even},\rightarrow}$ is of dimension $3\times 3$. Its first column is $\frac{1}{2}[A_0^{\mathcal{NN}}+A_0^{\mathcal{DN}}, A_1^{\mathcal{NN}}+A_1^{\mathcal{DN}}, A_2^{\mathcal{NN}}+A_2^{\mathcal{DN}}]^{\mathrm{T}}$, 
in which the coefficients $A_m^{\mathcal{NN}}$ and $A_m^{\mathcal{DN}}$ are computed for the case when the incident wave mode is $p=0$. The second and third columns are also of this form, but computed for the cases when $p=1$ and $p=2$, respectively. The matrix $S^{\text{odd},\leftarrow}_{\text{odd},\rightarrow}$ is of the dimension $2\times 2$. Its first column is $\tfrac{1}{2}[ A_1^{\mathcal{DD}}+A_1^{\mathcal{ND}}, A_2^{\mathcal{DD}}+A_2^{\mathcal{ND}}]^{\mathrm{T}}$, in which the coefficients $A_m^{\mathcal{DD}}$ and $A_m^{\mathcal{ND}}$ are computed for the case when the incident wave mode is $p=1$. The second column is of the same form, but calculated for the case when $p=2$.

\section{Time-domain results}
\label{time}
\noindent A time-domain solution is given as a continuous superposition of frequency domain solutions, as described in Eq.~\eqref{time_equation}. After approximating the continuous superposition using a quadrature rule, the computation of the time domain solution reduces to matrix multiplication \cite{Meylan20211,WILKS2025103421, wilks2024}. To do this, the integral over $k$ in \eqref{time_equation} is approximated using the quadrature rule of the form
\begin{equation}
\int_0^\infty g(k)\mathrm{d}k \approx\sum_{j=1}^{N_k} g(k_j)w_j,
\end{equation}
where $k_j=j\Delta k$ are the quadrature points and $w_{i}$ are the quadrature weights. In the case of the trapezoidal rule, these weights are
\begin{eqnarray*}
   w_{j}&=& \left\{
\begin{array}{ll}
    \Delta k/2 , & \textrm{if } j=1\text{ or }j=N_k,\\
     \Delta k, & \textrm{otherwise}.
   \end{array} 
\right.  
\end{eqnarray*}
After applying this quadrature rule, \eqref{time_equation} becomes
\begin{eqnarray}\label{td_sol_discrete}
 \varphi(x,y,t)&\approx&\mathrm{Re}\left\{ \sum_{j=1}^{N_k}  \hat{f}(k_{j})\phi (x,y,k_{j}) e^{-\upi k_{j} t}w_{j}\right\}.
 \label{mid}
\end{eqnarray}
Note that the general time domain solution is a superposition over incident modes $p$ and over all four incident directions, but for simplicity we restrict to the case where the incident mode $p$ is fixed and the wave is incident from $x=-\infty$. Next, we let $(x_{i},y_{i})$ for $1\leq j\leq N_x$ be the points at which we want to evaluate our time domain solution. Equation \eqref{td_sol_discrete} can then  be represented using matrix multiplication as
\begin{equation*}
\boldsymbol{\varphi}(t) 
\approx \mathrm{Re}\left\{[\phi]\mathrm{Diag}(w_je^{-\upi k_j t})\mathbf{\hat{f}}\right\}
\end{equation*}
where we have defined $\boldsymbol{\varphi}(t)$ to be the vector of length $N_x$ with entries $\varphi(x_i,y_i,t)$ and $[\phi]$ to be the matrix with entries $[\phi]_{ij}=\phi(x_i,y_i,k_j)$. Moreover, $\mathbf{\hat{f}}$ is the vector of length $N_k$ with entries $\hat{f}(k_j)$. Note that the calculation of the frequency domain solution matrix $[\phi]$ is the main computational expense of our time domain solution. In this formulation, this calculation only needs to be done once for a particular set of parameters, so that the solutions for different amplitude spectra $\hat{f}$ can be computed quickly.

\begin{figure}
    \centering
    \begin{subfigure}{0.48\textwidth}
        \includegraphics[height=5cm]{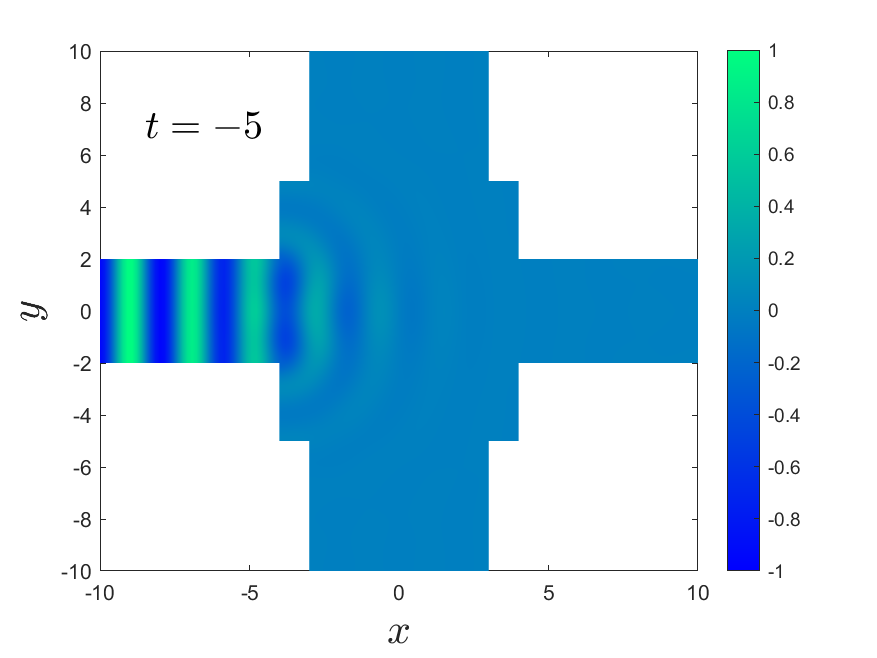}
        \caption{}
        \label{m1}
    \end{subfigure}
    \begin{subfigure}{0.48\textwidth}
        \includegraphics[height=5cm]{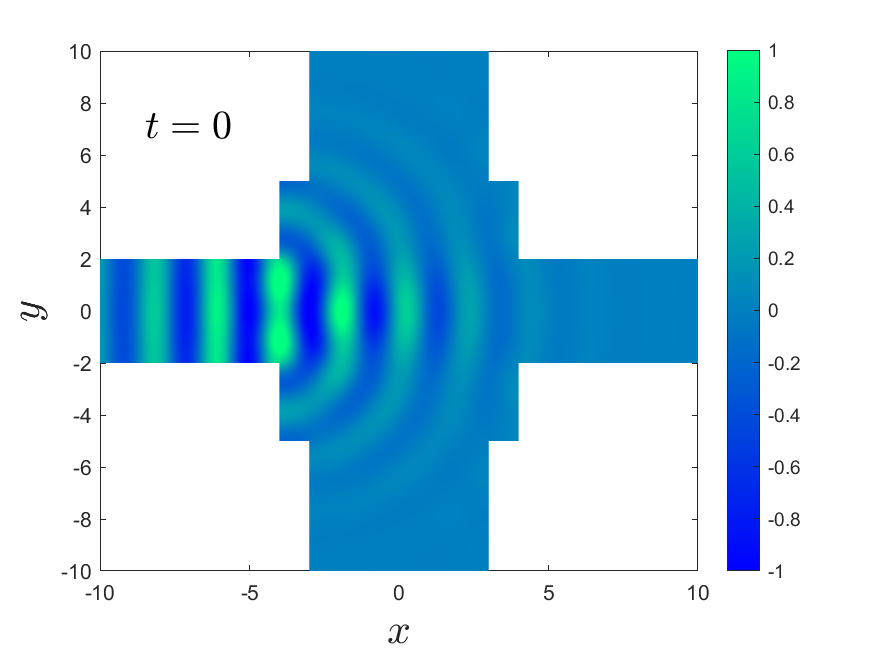}
        \caption{}
        \label{m2}
    \end{subfigure}
    
    \begin{subfigure}{0.48\textwidth}
        \includegraphics[height=5cm]{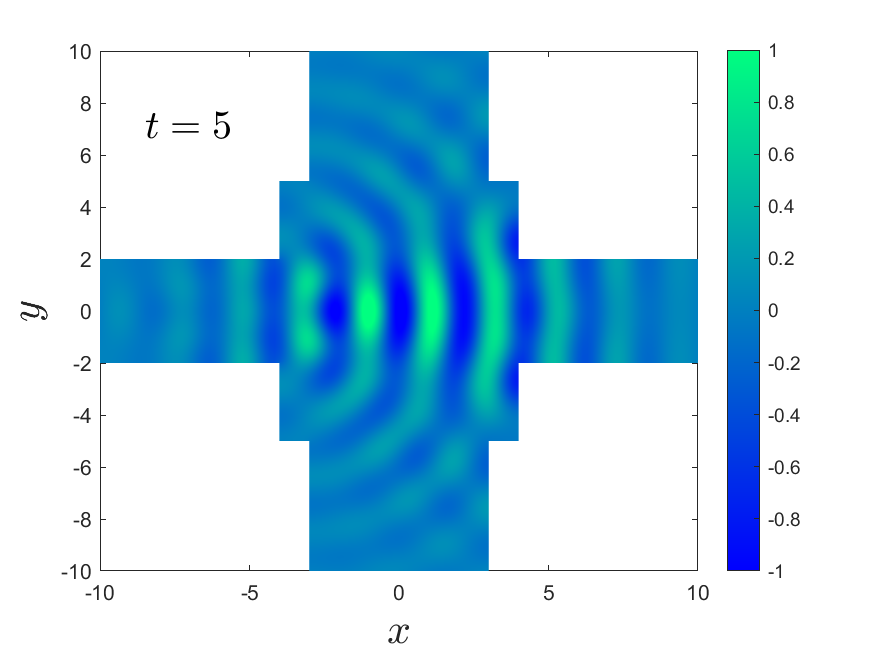}
        \caption{}
        \label{m3}
    \end{subfigure}
    \begin{subfigure}{0.48\textwidth}
        \includegraphics[height=5cm]{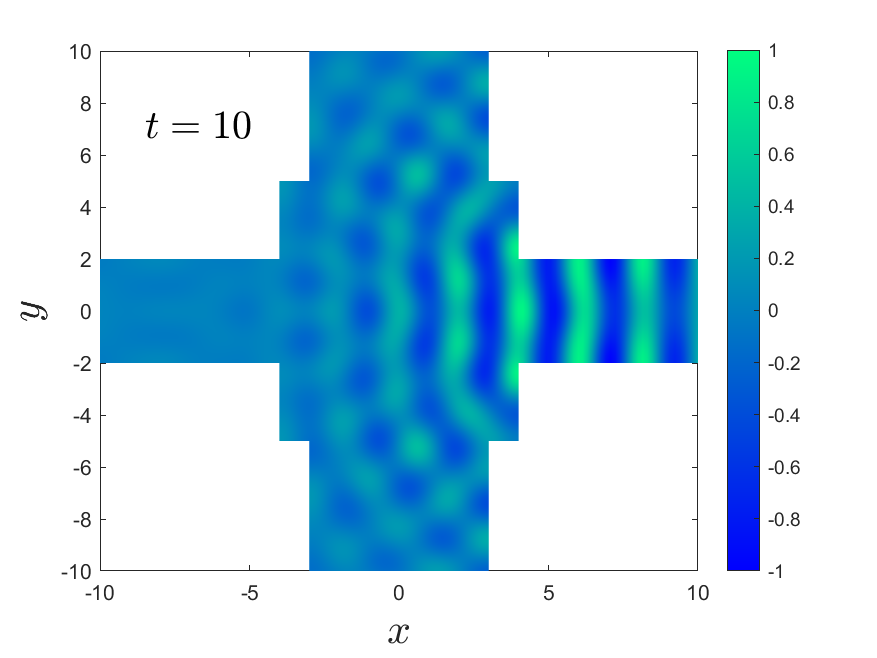}
        \caption{}
        \label{m4}
    \end{subfigure}
    
    \begin{subfigure}{0.48\textwidth}
        \includegraphics[height=5cm]{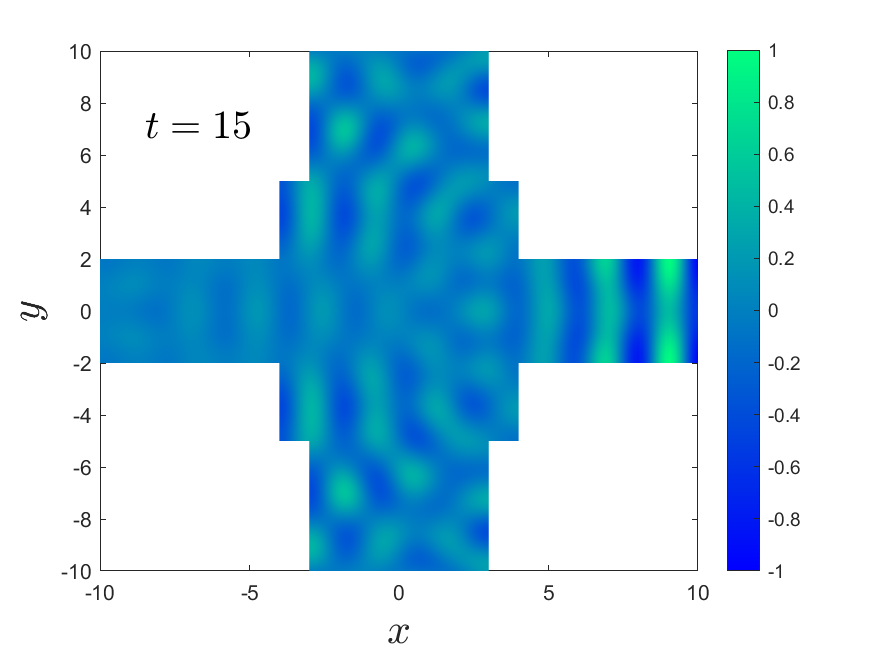}
        \caption{}
        \label{m5}
    \end{subfigure}
    \begin{subfigure}{0.48\textwidth}
        \includegraphics[height=5cm]{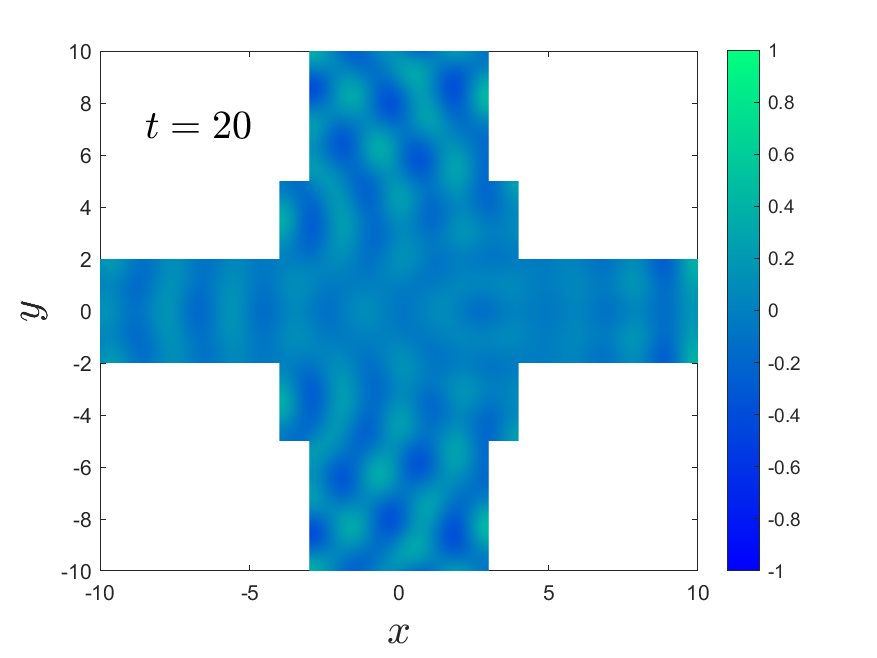}
        \caption{}
        \label{m6}
    \end{subfigure}
    
    \caption{Snapshots showing the scattering of a symmetric incident wave at times 
    (a) $t=-5$, (b) $t=0$, (c) $t=5$, (d) $t=10$, (e) $t=15$, and (f) $t=20$. 
    The parameters used are $p=0$, $b_1=5$, $b_2=4$, $a_1=2$, $a_2=3$, $k=10$, and $N=100$. 
    The incident spectrum was chosen to be $\hat{f}(k) = \frac{1}{\pi} e^{-8(k-3)^2}$.}
    \label{movie1}
\end{figure}

\begin{figure}
    \centering
    \begin{subfigure}{0.48\textwidth}
        \includegraphics[height=5cm]{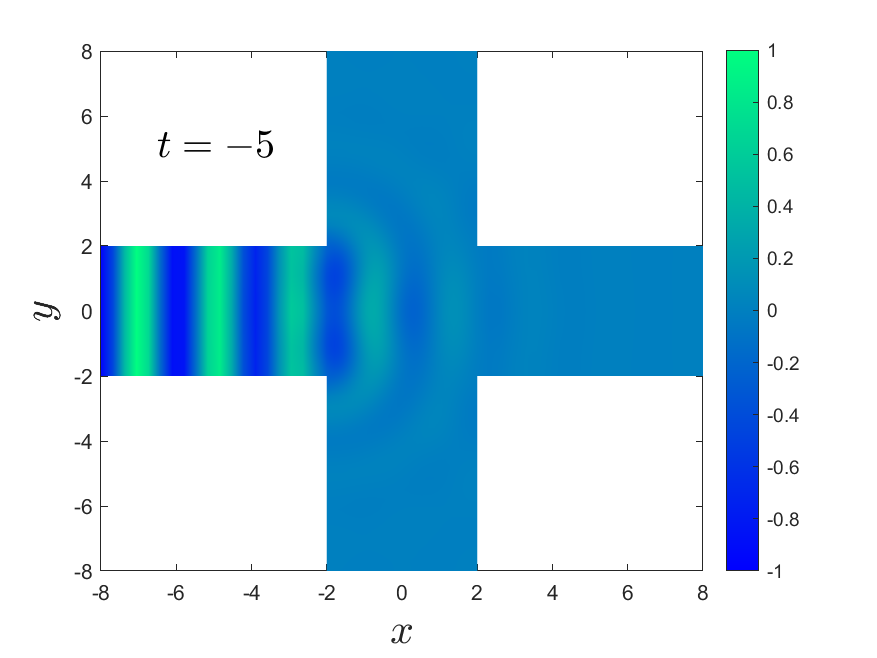}
        \caption{}
        \label{m1}
    \end{subfigure}
    \begin{subfigure}{0.48\textwidth}
        \includegraphics[height=5cm]{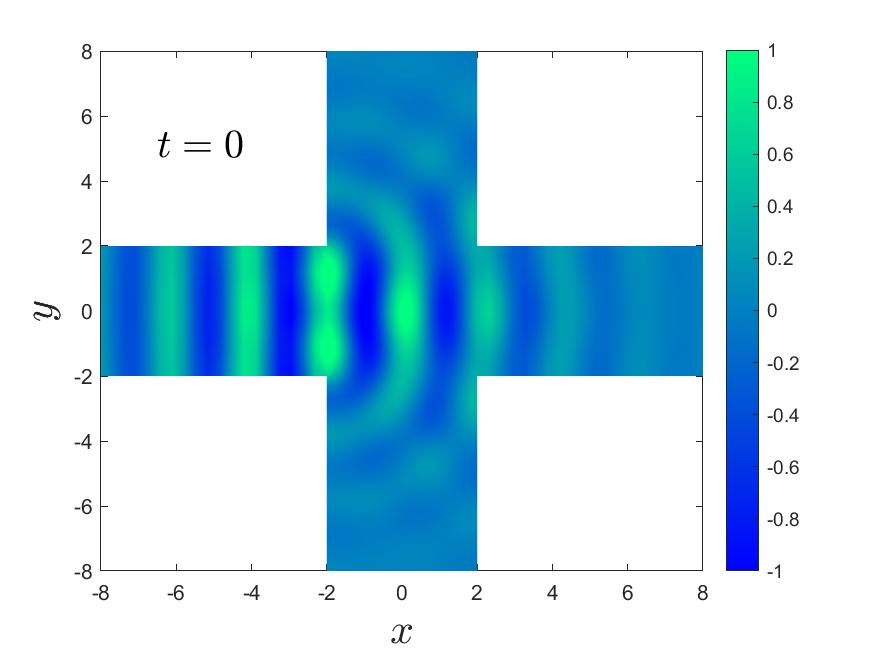}
        \caption{}
        \label{m2}
    \end{subfigure}
    \begin{subfigure}{0.48\textwidth}
        \includegraphics[height=5cm]{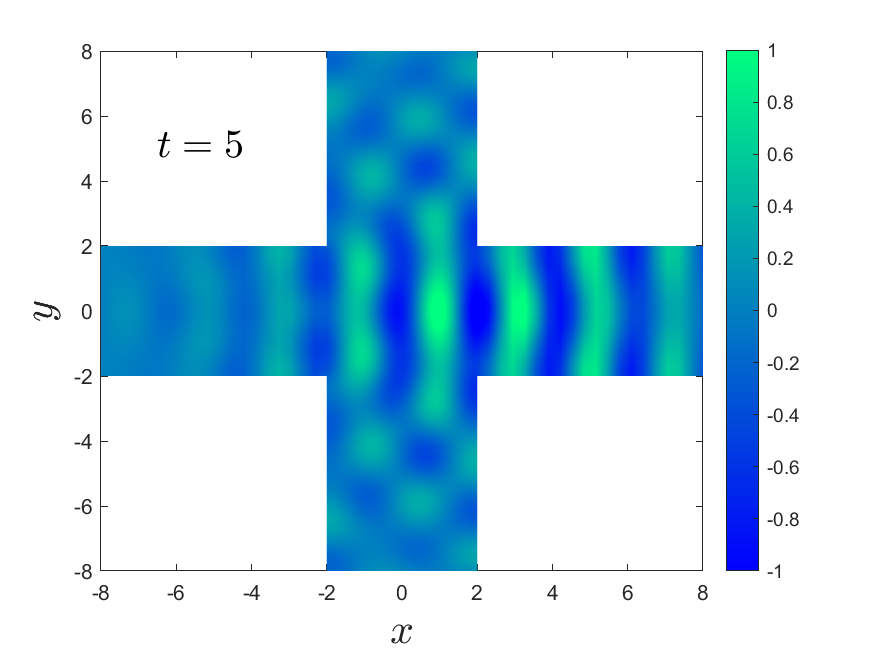}
        \caption{}
        \label{m3}
    \end{subfigure}
    \begin{subfigure}{0.48\textwidth}
        \includegraphics[height=5cm]{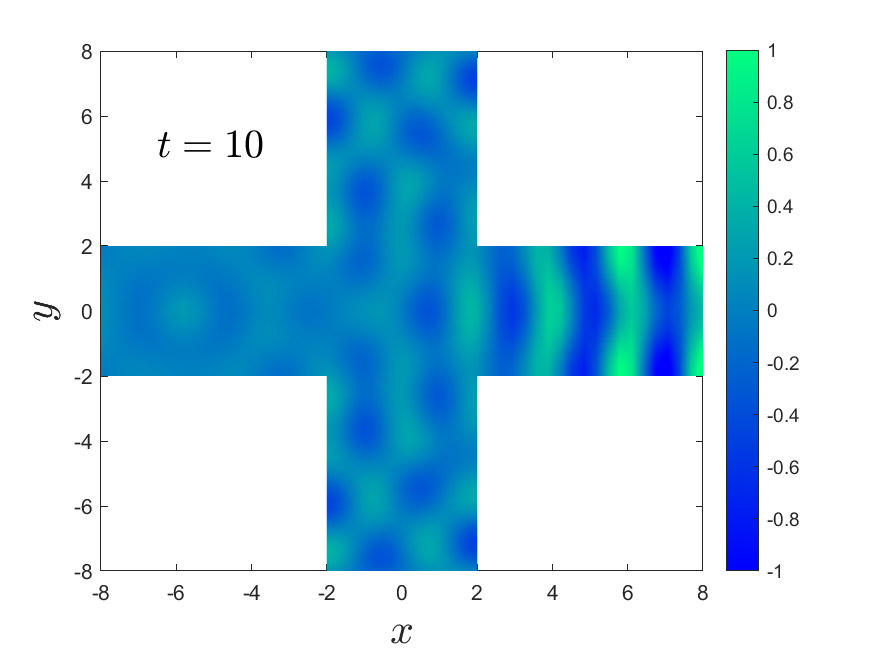}
        \caption{}
        \label{m4}
    \end{subfigure}
    \begin{subfigure}{0.48\textwidth}
        \includegraphics[height=5cm]{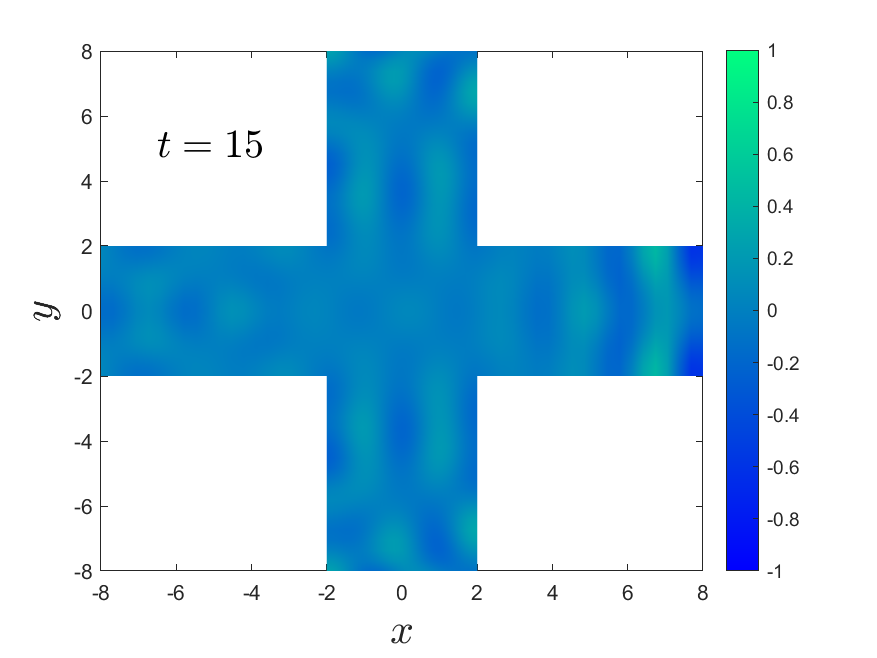}
        \caption{}
        \label{m5}
    \end{subfigure}
    \begin{subfigure}{0.48\textwidth}
        \includegraphics[height=5cm]{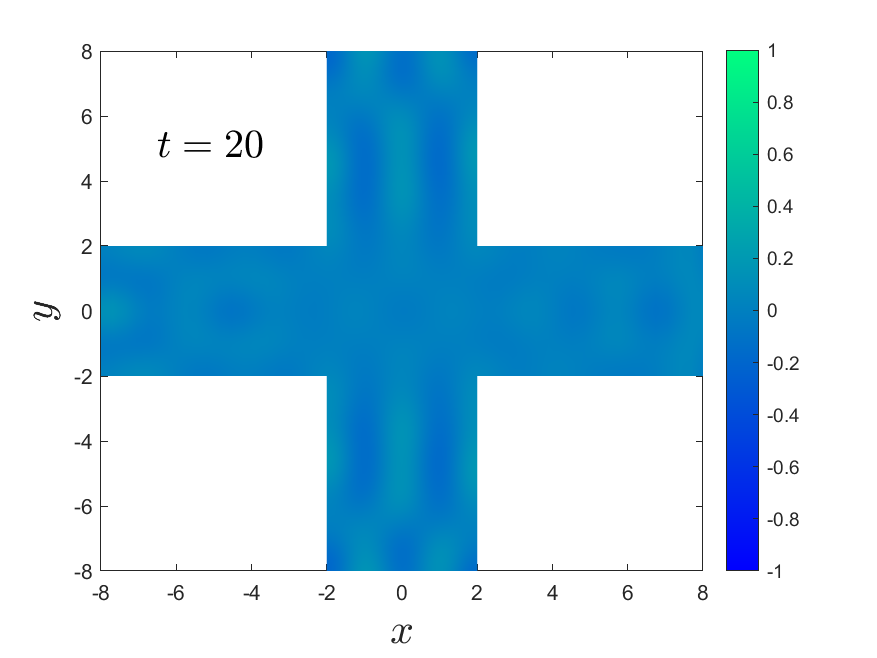}
        \caption{}
        \label{m6}
    \end{subfigure}
    \caption{Snapshots showing the scattering of a symmetric incident wave at times 
    (a) $t=-5$, (b) $t=0$, (c) $t=5$, (d) $t=10$, (e) $t=15$, and (f) $t=20$. 
    The parameters used are $p=0$, $N=100$, $k=15$, and $a_1=a_2=b_1=b_2=2$. 
    The incident spectrum was chosen to be $\hat{f}(k) = \frac{1}{\pi} e^{-8(k-3)^2}$.}
    \label{movie2}
\end{figure}

\begin{figure}
    \centering
    \begin{subfigure}{0.48\textwidth} 
        \includegraphics[height=4.8cm]{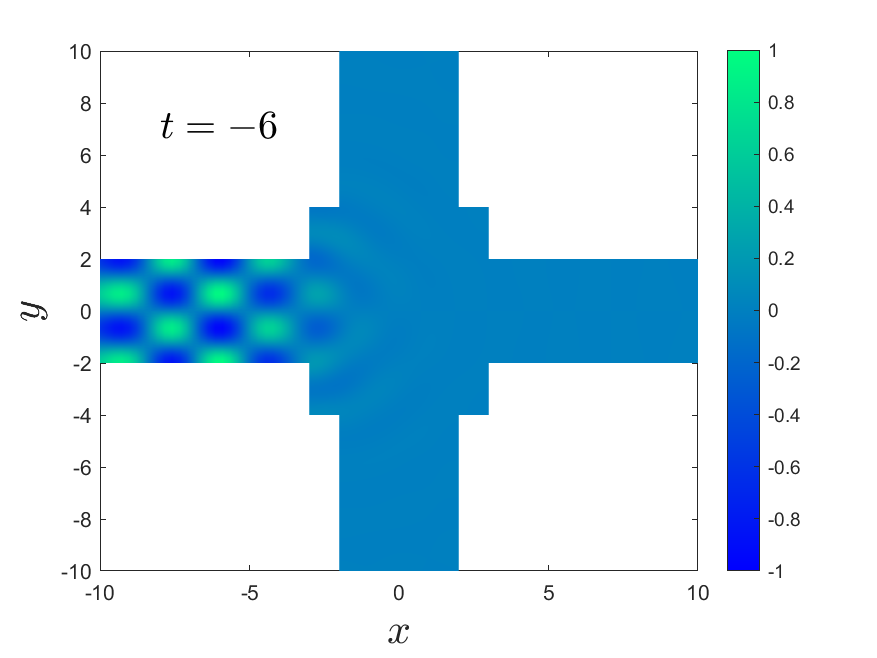}
        \caption{}
        \label{mm1}
    \end{subfigure}
    \hspace{0.02\textwidth} 
    \begin{subfigure}{0.48\textwidth}
        \includegraphics[height=4.8cm]{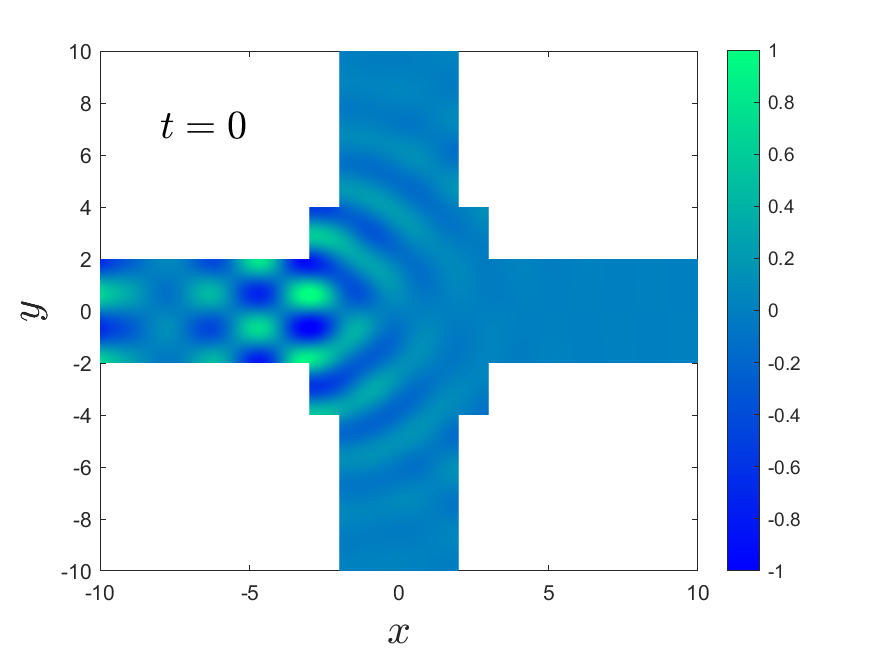}
        \caption{}
        \label{mm2}
    \end{subfigure}\\[0.5cm] 
    \begin{subfigure}{0.48\textwidth}
        \includegraphics[height=4.8cm]{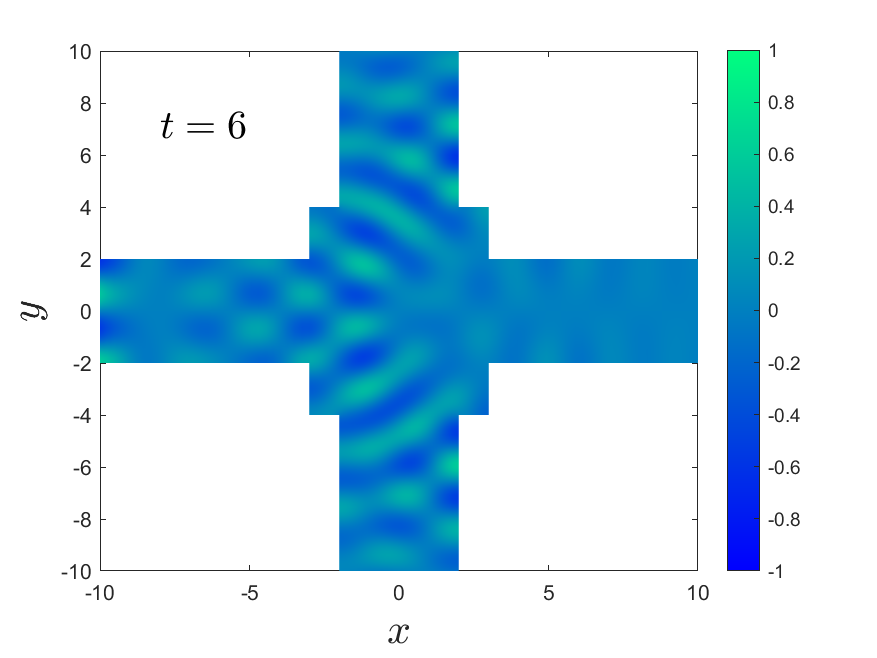}
        \caption{}
        \label{mm3}
    \end{subfigure}
    \hspace{0.02\textwidth}
    \begin{subfigure}{0.48\textwidth}
        \includegraphics[height=4.8cm]{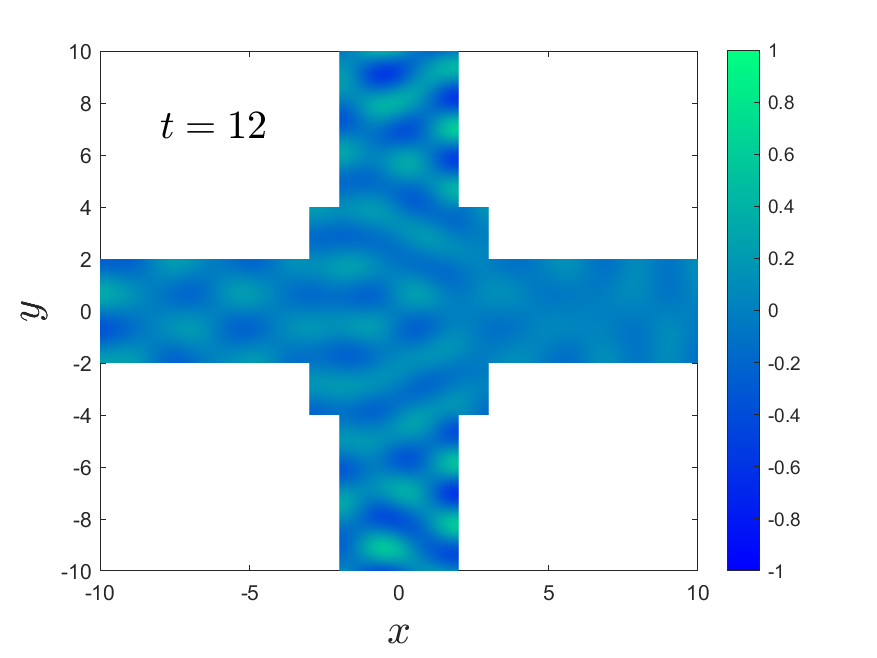}
        \caption{}
        \label{mm4}
    \end{subfigure}\\[0.5cm]
    \begin{subfigure}{0.48\textwidth}
        \includegraphics[height=4.8cm]{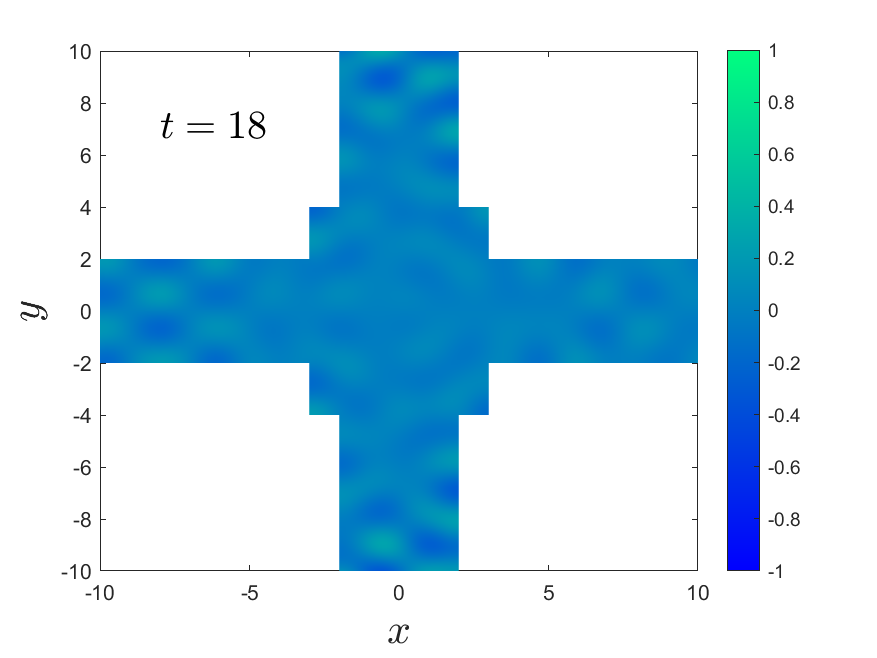}
        \caption{}
        \label{mm5}
    \end{subfigure}
    \hspace{0.02\textwidth}
    \begin{subfigure}{0.48\textwidth}
        \includegraphics[height=4.8cm]{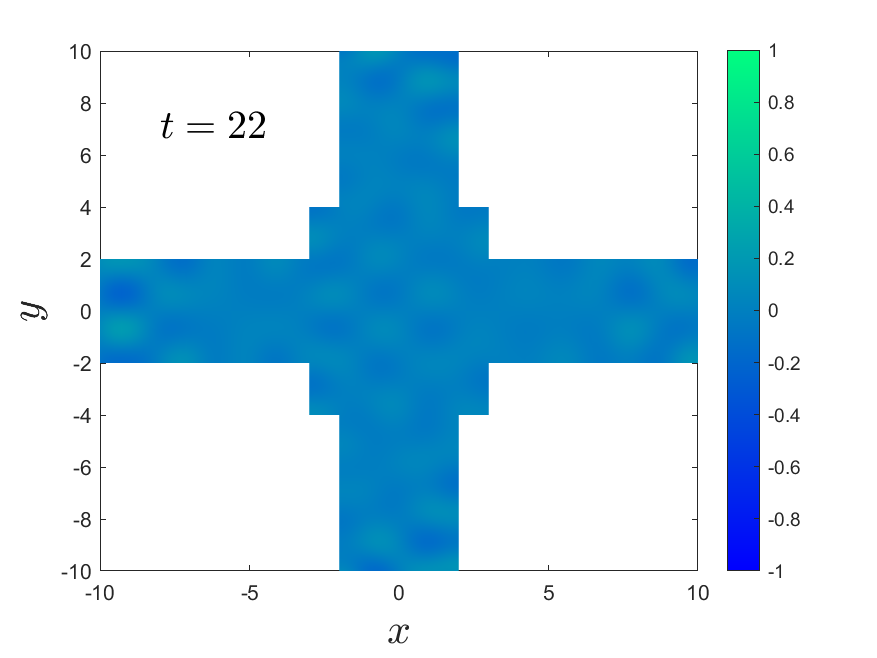}
        \caption{}
        \label{mm6}
    \end{subfigure}
    \caption{
        Snapshots showing the scattering of an antisymmetric incident wave at times: 
        (a) $t=-6$, (b) $t=0$, (c) $t=6$, (d) $t=12$, (e) $t=18$, and (f) $t=22$.\\
        The parameters used are $p=1$, $b_1=4$, $b_2=3$, $a_1 = a_2 = 2$, $k=12$, and $N=100$. 
        The incident spectrum is $\hat{f}(k) = \frac{1}{\pi} e^{-9(k-3)^2}$.
    }
    \label{movie3}
\end{figure}

\begin{figure}
    \centering
    \begin{subfigure}{0.48\textwidth} 
        \includegraphics[height=4.8cm]{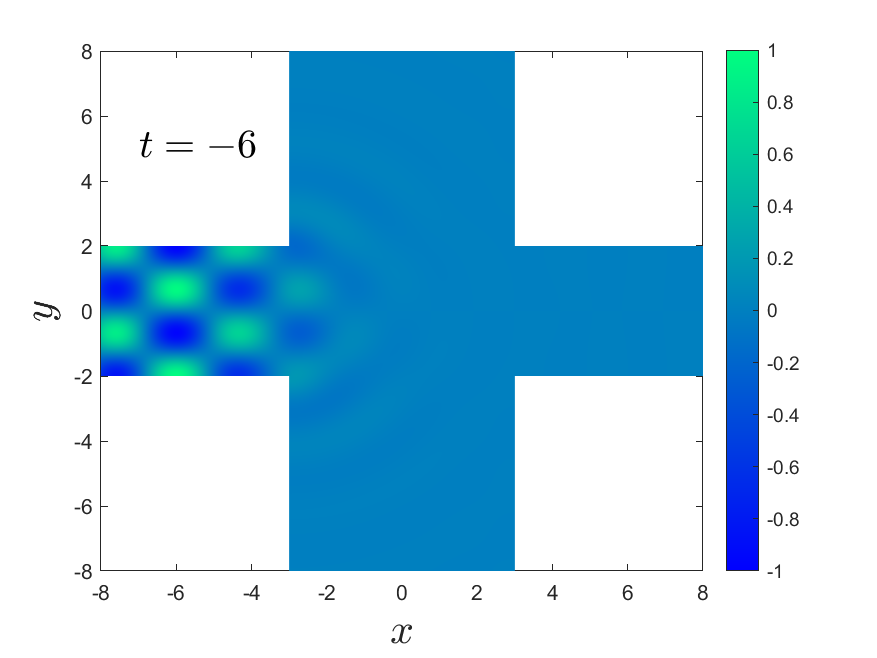} 
        \caption{}
        \label{r1}
    \end{subfigure}
    \hspace{0.02\textwidth} 
    \begin{subfigure}{0.48\textwidth}
        \includegraphics[height=4.8cm]{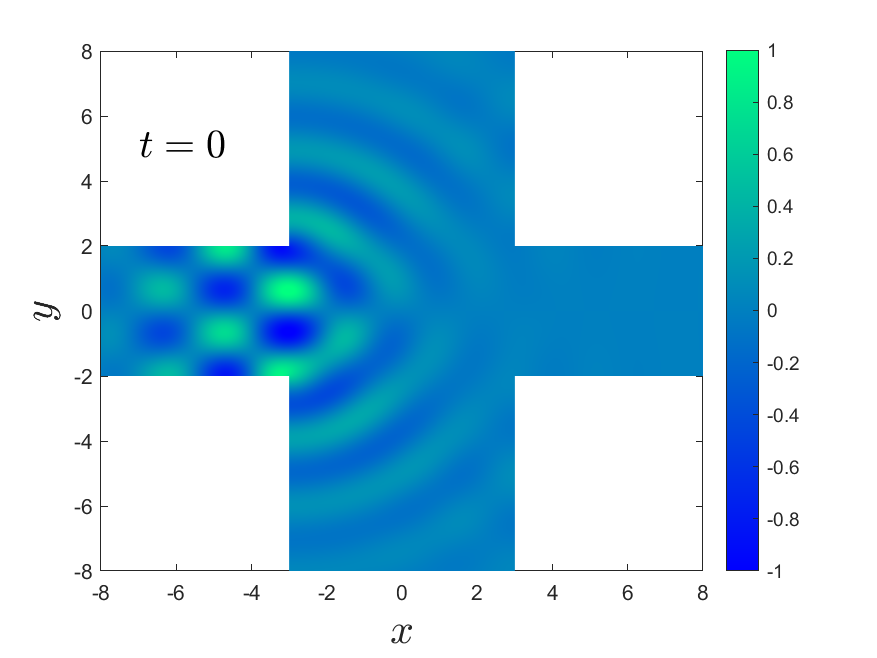}
        \caption{}
        \label{r2}
    \end{subfigure}\\[0.5cm] 
    \begin{subfigure}{0.48\textwidth}
        \includegraphics[height=4.8cm]{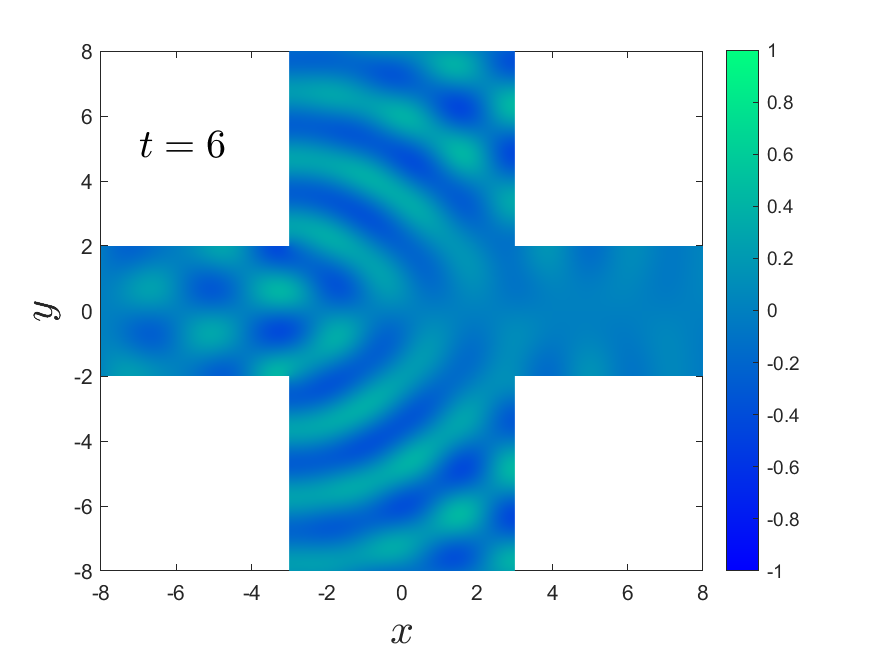}
        \caption{}
        \label{r3}
    \end{subfigure}
    \hspace{0.02\textwidth}
    \begin{subfigure}{0.48\textwidth}
        \includegraphics[height=4.8cm]{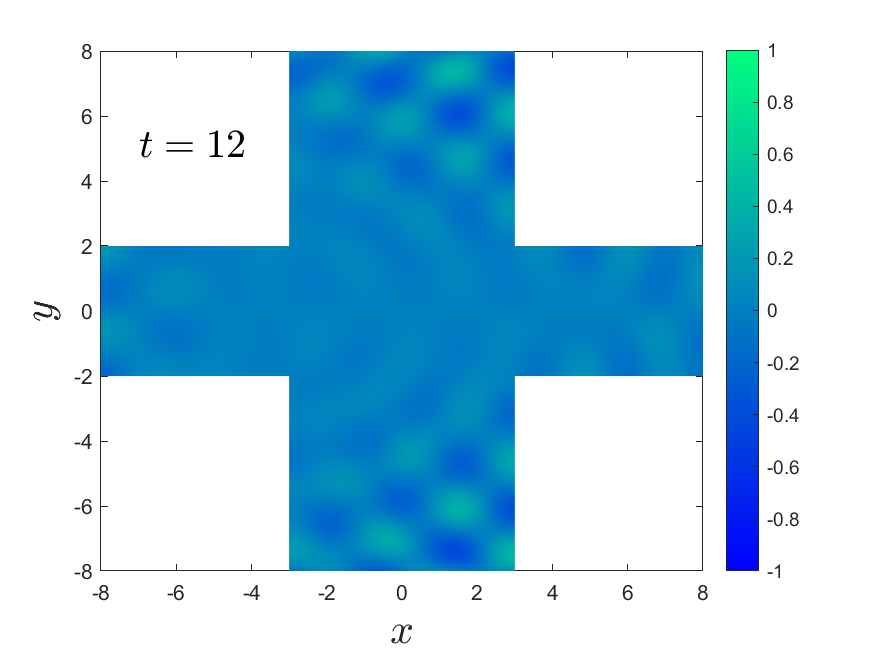}
        \caption{}
        \label{r4}
    \end{subfigure}\\[0.5cm]
    \begin{subfigure}{0.48\textwidth}
        \includegraphics[height=4.8cm]{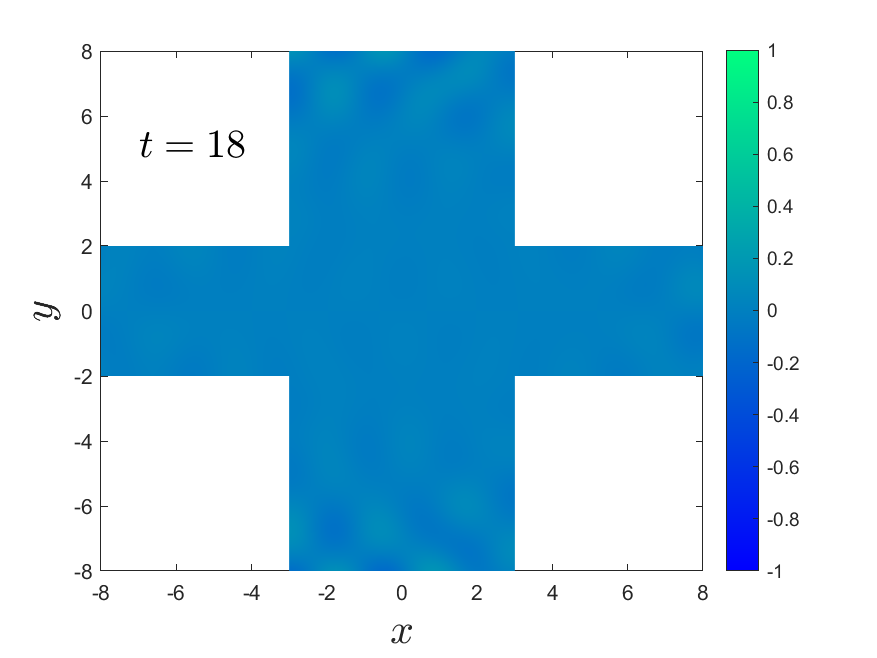}
        \caption{}
        \label{r5}
    \end{subfigure}
    \hspace{0.02\textwidth}
    \begin{subfigure}{0.48\textwidth}
        \includegraphics[height=4.8cm]{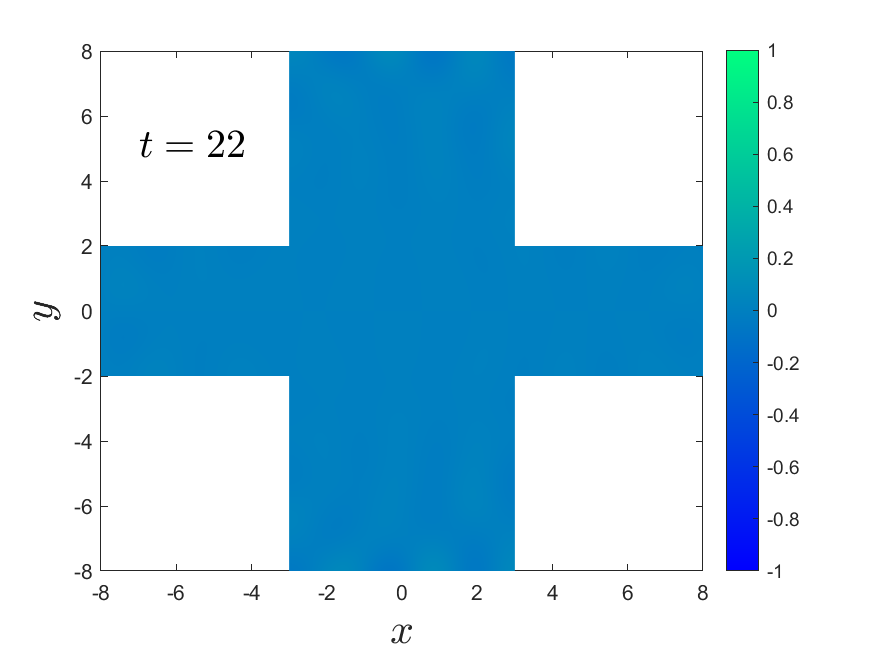}
        \caption{}
        \label{r6}
    \end{subfigure}
    \caption{
        Showing the scattering of an antisymmetric incident wave at times: 
        (a) $t=-6$, (b) $t=0$, (c) $t=6$, (d) $t=12$, (e) $t=18$, and (f) $t=22$.\\
        The parameters used are $p=1$, $b_1=4$, $b_2=a_2=3$, $a_2$, $k=8$, and $N=100$. 
        The incident spectrum was chosen as $\hat{f}(k) = \frac{1}{\pi} e^{-9(k-3)^2}$.
    }
    \label{movie4}
\end{figure}

Figures ~\ref{movie1}--\ref{movie4} show time domain solutions for a range of different parameters and incident spectra $\hat{f}$. In each figure, the panels show snapshots from Movies 1--4 at a range of different times, which illustrate the scattering of an incident wave travelling from left to right. The special cases $a_1=b_1$ and $a_2=b_2$, in which the dimensions of the rectangular region match those of the waveguides, are shown in Figures \ref{movie2} and \ref{movie4}. Figure ~\ref{movie3} shows the case where the incident wave is of odd symmetry about $y=0$, with the remaining figures showing cases of even incident waves. Note that all Movies are provided in the supplementary material.

\section{Conclusions}
\label{conclusions}
\noindent In this investigation, we applied a symmetry-based decomposition and the eigenfunction matching technique to solve the problem of wave scattering at a rectangular junction of four waveguides in the frequency domain. Subsequently, the time-domain solution was obtained as a continuous superposition of frequency domain solutions, and results were given for a range of incident pulses and waveguide geometries. Additionally, we described how to use our solution to obtain the scattering matrix for the junction. Future work could use this scattering matrix to solve the problem of wave propagation through a square lattice of such junctions, thereby forming a connection with the subresonant lattices studied in quantum graph theory \cite{Lawrie2022,Lawrie2023}

\appendix

\section{Eigenfunction matching of the remaining cases}
\label{matchingnew}
\noindent We determine the matching conditions of both discontinuities in the case of Dirichlet-Dirichlet problem. On using \eqref{incident2}, \eqref{solutionanti1} and \eqref{solutionanti2} into the matching condition \eqref{matching1} to obtain 
\begin{eqnarray}
\widetilde \Phi_p(y)+\sum_{n=0}^{\infty} A^{\mathcal{DD}}_{n}  \widetilde \Phi_n(y)&=& \sum_{n=0}^{\infty} B^{\mathcal{DD}}_{n}  \widetilde\Gamma_n(y)\\
  &+&\sum_{n=0}^{\infty} C^{\mathcal{DD}}_{n} \left(\frac{\sinh(i\bar\zeta_{n}y)}{\sinh(i\bar\zeta_{n}b_1)} \right) \sin\left( (-2n+1)\frac{\pi}{2}\right)\nonumber .   
\end{eqnarray}
Taking the inner product with $\overline\Phi_m(y)$ and integrating over $[0,a_1]$, then using Eq.~\eqref{hat u}, yields
\begin{equation}
     \hat I_{mp}(a_1)+ A^{\mathcal{DD}}_{m} \hat I_{mn}(a_1)=\sum_{n=0}^{\infty} B^{\mathcal{DD}}_{n} R_{mn}+\sum_{n=0}^{\infty} C^{\mathcal{DD}}_{n} Q_{mn}, \quad m=0,1,2,\dots,
      \label{system5}
\end{equation}
where $R_{mn}$ and $Q_{mn}$ are stated in \ref{R} and \ref{Q}. On using \eqref{soltuionant3}, \eqref{solutionanti1} and \eqref{solutionanti2} into  \eqref{matching2} to derive the matching at $y=b_1$ as 
 \begin{equation}
        \sum_{n=0}^{\infty} D^{\mathcal{DD}}_n \widetilde \aleph_{n}(x)=\sum_{n=0}^{\infty} B^{\mathcal{DD}}_{n} \left( \frac{\sinh(i\bar\lambda_{n}x)}{\sinh(-i\bar\lambda_{n}b_{2})}\right)\sin((2n+1)\pi/2)+\sum_{n=0}^{\infty} C^{\mathcal{DD}}_{n} \widetilde\Upsilon_n(x),
 \end{equation}
we take the inner product with $\overline\Omega_{m}(x)$, and integrate over $[-a_2,0]$, yields 
\begin{equation}
     D^{\mathcal{DD}}_{m} \hat I_{mn}(a_2)=\sum_{n=0}^{\infty} B^{\mathcal{DD}}_{n} W_{mn}+ \sum_{n=0}^{\infty} C^{\mathcal{DD}}_{n} K_{mn} , \quad m=0,1,2,\dots,
       \label{system6}
\end{equation}

\noindent where $W_{mn}$ and $K_{mn}$ are stated in \ref{W} and \ref{K}. In addition, on using the derivative of \eqref{incident2}, \eqref{solutionanti1} and \eqref{solutionanti2} with respect to $x$ into the matching condition \eqref{dreivativex} to obtain 
\begin{equation}
    \bar \gamma_p \widetilde \Phi_p(y)- \sum_{n=0}^{\infty} A^{\mathcal{DD}}_{n} \bar\gamma_{n} \widetilde \Phi_n(y) =\sum_{n=0}^{\infty} B^{\mathcal{DD}}_{n} \bar\lambda_{n} \coth(-i\bar\lambda_{n}b_{2} \widetilde\Gamma_n(y).
\end{equation}
Taking the inner product with $\widetilde\Gamma_n(y)$, and integrate over $[0,b_1]$ and then using Eq.~\eqref{R}, yields  
 \begin{equation}
      \bar\gamma_p R_{pm}- \sum_{n=0}^{\infty} A^{\mathcal{DD}}_{n} \bar\gamma_n R_{nm}=  B^{\mathcal{DD}}_{m}  \bar\lambda_{m} \coth(-i\bar\lambda_{m}b_{2}) \hat I_{mn}(b_1), \quad m=0,1,2,\dots.
        \label{system7}
  \end{equation}
On using the derivative of \eqref{soltuionant3}, \eqref{solutionanti1} and \eqref{solutionanti2} with respect to $y$ into \eqref{dreivativey} to obtain 
\begin{equation}
    \sum_{n=0}^{\infty} D^{\mathcal{DD}}_{n} \bar\zeta^{\prime}_{n} \widetilde\aleph_{n}(x)= \sum_{n=0}^{\infty} C^{\mathcal{DD}}_{n} \bar\zeta_{n}  \coth(i\bar\zeta_{n}b_{1})\widetilde \Upsilon_{n}(x).
\end{equation}
Taking the inner product with $\widetilde\Upsilon_m(y)$ and integrate over $[-b_2,0]$ and then using \eqref{K}, yields 
\begin{eqnarray}
  \sum_{n=0}^{\infty} D^{\mathcal{DD}}_{n}\bar\zeta^{\prime}_{n} K_{nm}  &=&C^{\mathcal{DD}}_{m}  \bar\zeta_{m}  \coth(i\bar\zeta_{m}b_{1}) \hat I_{mn}(b_2), \quad m=0,1,2,\dots.
    \label{system8}
\end{eqnarray}

\noindent We proceeded with an almost identical method to the previous two cases to obtain the matching conditions in the Neumann-Dirichlet problem. We used \eqref{incident2}, \eqref{mix1} and \eqref{mix2} and then \eqref{solution31}, \eqref{mix1} and \eqref{mix2} are matched using \eqref{matching1} and \eqref{matching2} respectively. Then, using \eqref{hat u}, \eqref{R}, \eqref{E} to obtain the expressions 

\begin{eqnarray}
\label{system11}
 \hat I_{mp}(a_1)+ A^{\mathcal{ND}}_{m} \hat I_{mn}(a_1)&=&\sum_{n=0}^{\infty} B^{\mathcal{ND}}_{n} R_{mn}+\sum_{n=0}^{\infty} C^{\mathcal{ND}}_{n} V^{\prime}_{mn}, \quad m=0,1,2,\dots,\\
    D^{\mathcal{ND}}_{m} I_{mn}(a_1)&=&\sum_{n=0}^{\infty} B^{\mathcal{ND}}_{n} W^{\prime}_{mn}+\sum_{n=0}^{\infty} C^{\mathcal{ND}}_{n} E_{mn}, \quad m=0,1,2,\dots,   
    \label{system12}
\end{eqnarray}

where $V^{\prime}_{mn}$ and $W^{\prime}_{mn}$ are given in \ref{Vhat} and \ref{What}. The derivative of \eqref{incident2}, \eqref{mix1} and \eqref{mix2} and then \eqref{solution31}, \eqref{mix1} and \eqref{mix2} with respect to $x$ and then $y$ are matched using \eqref{dreivativex} and \eqref{dreivativey} respectively. Simplifying by using \eqref{R} and \eqref{E} to obtain 
\begin{eqnarray}
\label{system13}
    \bar\gamma_p R_{pm}- \sum_{n=0}^{\infty} A^{\mathcal{ND}}_{n} \bar\gamma_n R_{nm}&=& B^{\mathcal{ND}}_{m}  \bar\lambda_{m} \tanh(-i\bar\lambda_{m}b_{2}) \hat I_{mn}(b_1), \quad m=0,1,2,\dots, \\
     \sum_{n=0}^{\infty} D^{\mathcal{ND}}_{n} \bar\eta_{n}  E_{nm} &=& C^{\mathcal{ND}}_{m} \bar\eta^{\prime}_{m} \coth(i\bar\eta^{\prime}_{m}b_{1}) I_{mn}(b_2), \quad m=0,1,2,\dots, 
     \label{system14}
\end{eqnarray}
Now, we derive the matching conditions in the Dirichlet-Neumann problem and follow a nearly identical process. We match \eqref{incident}, \eqref{mix3} and \eqref{mix4} into \eqref{matching1} and then \eqref{soltuionant3}, \eqref{mix3} and \eqref{mix4} into ~\eqref{matching2}. Using \eqref{H} and \eqref{K} to obtain 
\begin{eqnarray}
\label{system21}
     I_{mp}(a_1)+ A^{\mathcal{DN}}_m  I_{mn}(a_1)&=&\sum_{n=0}^{\infty} B^{\mathcal{DN}}_{n} H_{mn}+ \sum_{n=0}^{\infty} C^{\mathcal{DN}}_{n} Q^{\prime}_{mn}, \quad m=0,1,2,\dots, \\
       D^{\mathcal{DN}}_{m} \hat I_{mn}(a_2)&=& \sum_{n=0}^{\infty} B^{\mathcal{DN}}_{n}  O^{\prime}_{mn}+ \sum_{n=0}^{\infty} C^{\mathcal{DN}}_{n} K_{mn}, \quad m=0,1,2,\dots,
       \label{system22}
\end{eqnarray}
where $Q^{'}_{mn}$ and $O^{'}_{mn}$ are given in \ref{Qhat} and \ref{Ohat}. Then, we substitute the derivative of \eqref{incident}, \eqref{mix3} and \eqref{mix4} into \eqref{dreivativex} and then \eqref{soltuionant3}, \eqref{mix3} and \eqref{mix4} into \eqref{dreivativey} with respect to $x$ and then $y$. Using \eqref{H} and \eqref{K} to obtain 
\begin{eqnarray}
\label{system23}
      \bar \beta_p H_{pm}- \sum_{n=0}^{\infty} A^{\mathcal{DN}}_{n} \bar\beta_{n} H_{nm}&=&  B^{\mathcal{DN}}_{m} \bar\alpha_{m} \coth(-i\bar\alpha_{m}b_{2}) I_{mn}(b_1),\\
        \sum_{n=0}^{\infty} D^{\mathcal{DN}}_{n} \bar\zeta^{\prime}_{n}  K_{nm} &=& C^{\mathcal{DN}}_{m} \bar\zeta_{m}\tanh(i\bar\zeta_{m}b_{1}) \hat I_{mn}(a_2).
        \label{system24}
\end{eqnarray}

\noindent To solve these systems numerically, we express the system of equations \eqref{system5}, \eqref{system6}, \eqref{system7} and \eqref{system8} as

{\footnotesize
\begin{equation*}
\begin{array}{c}
    \begin{bmatrix}
        \left[-\textrm{diag}(\hat{I}_{mn}(a_1))\right] & \left[R_{mn}\right] & \left[Q_{mn}\right] & \left[0\right] \\
        \left[0\right] & \left[W_{mn}\right] & \left[K_{mn}\right] & \left[-\textrm{diag}(\hat{I}_{mn}(a_2))\right] \\
        \left[\textrm{diag}(\bar{\gamma}_{n}) R^{T}_{mn}\right] & \left[\textrm{diag}(\bar{\lambda}_{m} \coth(-i\bar{\lambda}_{m} b_{2}) \hat{I}_{mn}(b_1))\right] & \left[0\right] & \left[0\right] \\
        \left[0\right] & \left[0\right] & \left[\textrm{diag}(\bar{\zeta}_{m} \coth(i\bar{\zeta}_{m} b_{1}) \hat{I}_{mn}(b_{2}))\right] & \left[-\textrm{diag}(\bar{\zeta}'_{n}) K^{T}_{mn}\right]
    \end{bmatrix}\\
    \begin{bmatrix}
        \left[A^{\mathcal{DD}}_n\right] \\
        \left[B^{\mathcal{DD}}_n\right] \\
        \left[C^{\mathcal{DD}}_n\right] \\
        \left[D^{\mathcal{DD}}_n\right]
    \end{bmatrix}
    =
    \begin{bmatrix}
        \left[\hat{I}_{mn}(a_1)\right] \\
        \left[0\right] \\
        \left[\bar{\gamma}_{n} R^{T}_{pm}\right] \\
        \left[0\right]
    \end{bmatrix}.
\end{array}
\end{equation*}}

\noindent Additionally, we express \eqref{system11}, \eqref{system12}, \eqref{system13} and \eqref{system14} as

{\footnotesize
\begin{equation*}
\begin{array}{c}
    \begin{bmatrix}
        \left[-\textrm{diag}(\hat{I}_{mn}(a_1))\right] & \left[R_{mn}\right] & \left[V'_{mn}\right] & \left[0\right] \\
        \left[0\right] & \left[W'_{mn}\right] & \left[E_{mn}\right] & \left[-\textrm{diag}(\hat{I}_{mn}(a_2))\right] \\
        \left[\textrm{diag}(\bar{\gamma}_{n}) R^{T}_{mn}\right] & \left[\textrm{diag}(\bar{\lambda}_{m} \tanh(-i\bar{\lambda}_{m}b_{2}) \hat{I}_{mn}(b_1))\right] & \left[0\right] & \left[0\right] \\
        \left[0\right] & \left[0\right] & \left[\textrm{diag}(\bar{\eta}'_{m} \coth(i\bar{\eta}'_{m}b_{1}) \hat{I}_{mn}(b_2))\right] & \left[-\textrm{diag}(\bar{\eta}_{n}) E^{T}_{mn}\right]
    \end{bmatrix}\\
    \begin{bmatrix}
        \left[A^{\mathcal{ND}}_n\right] \\
        \left[B^{\mathcal{ND}}_n\right] \\
        \left[C^{\mathcal{ND}}_n\right] \\
        \left[D^{\mathcal{ND}}_n\right]
    \end{bmatrix}
    =
    \begin{bmatrix}
        \left[\hat{I}_{mn}(a_1)\right] \\
        \left[0\right] \\
        \left[\bar{\gamma}_{n} R^{T}_{pm}\right] \\
        \left[0\right]
    \end{bmatrix}.
\end{array}
\end{equation*}}

\noindent Then, we express \eqref{system21}, \eqref{system22}, \eqref{system23} and \eqref{system24} as

{\footnotesize
\begin{equation*}
\begin{array}{c}
    \begin{bmatrix}
        \left[-\textrm{diag}(I_{mn}(a_1))\right] & \left[H_{mn}\right] & \left[Q'_{mn}\right] & \left[0\right] \\
        \left[0\right] & \left[O'_{mn}\right] & \left[K_{mn}\right] & \left[-\textrm{diag}(\hat{I}_{mn}(a_2))\right] \\
        \left[\textrm{diag}(\bar{\beta}_{n}) H^{T}_{mn}\right] & \left[\textrm{diag}(\bar{\alpha}_{m} \coth(-i\bar{\alpha}_{m}b_{2}) I_{mn}(b_1))\right] & \left[0\right] & \left[0\right] \\
        \left[0\right] & \left[0\right] & \left[\textrm{diag}(\bar{\zeta}_{m} \tanh(i\bar{\zeta}_{m}b_{1}) \hat{I}_{mn}(b_2))\right] & \left[-\textrm{diag}(\bar{\zeta}'_{n}) K^{T}_{mn}\right]
    \end{bmatrix}\\
    \begin{bmatrix}
        \left[A^{\mathcal{DN}}_n\right] \\
        \left[B^{\mathcal{DN}}_n\right] \\
        \left[C^{\mathcal{DN}}_n\right] \\
        \left[D^{\mathcal{DN}}_n\right]
    \end{bmatrix}
    =
    \begin{bmatrix}
        \left[I_{mn}(a_1)\right] \\
        \left[0\right] \\
        \left[\bar{\beta}_{n} H^{T}_{pm}\right] \\
        \left[0\right]
    \end{bmatrix}.
\end{array}
\end{equation*}}
Where the outer bracket denotes the matrix.  

\section{The matrix elements calculations}
\label{Calculations}

\noindent We explore the integrations that have been used to solve the four matrices, respectively   

{\footnotesize
\begin{eqnarray}
  \label{H}
   H_{mn}&=& \int_{0}^{a_1}  \Gamma_n(y) \Phi_m(y) dy \\
    &=& \left\{
\begin{array}{ll}  
a_1, & \textrm{if}\ \beta_m = \alpha_n=0,\\
a_1/2 + \sin(2 a_1\beta_m)/(4\beta_m), & \textrm{if}\ \beta_m =\alpha_n\neq 0,\\
  \sin(a_1 (\alpha_n + \beta_m))/(2 (\alpha_n + \beta_m)) \\
  + \sin(a_1 (\alpha_n - \beta_m))/(2 (\alpha_n - \beta_m)), & \textrm{if} \ \beta_m \neq \alpha_n \nonumber.
\end{array} 
\right.  \\ 
\label{V}
 V_{mn}&=& \int_{0}^{a_1}  \left(\cosh(i\bar  \eta^{'}_n y)/\cosh(i\bar  \eta^{'}_n b_1) \right) \cos(-n \pi) \Phi_m(y) dy \\
    &=& \left\{
\begin{array}{ll}  
 a_1/\cosh(i\bar\eta^{'}_n b_1) \cos(-n \pi), & \textrm{if}\ \beta_m = \bar\eta_n^{'}=0,\\
\sin(a_1\beta_m (1 - 1i)) (1 + 1i) + \sin(a_1\beta_m (1 + 1i))\\
(1 - 1i))/(4\beta_m)
/\cosh(i\bar\eta^{'}_n b_1) \cos(-n \pi), & \textrm{if}\ \beta_m = \bar\eta_n^{'}\neq 0,\\
   (\beta_m\cos(a_1\eta_n)\sin(a_1\beta_m) - \bar\eta^{'}_n\cos(a_1\beta_m)\sin(a_1\bar\eta^{'}_n))\\
   /(\beta_m^2 - \bar\eta^{2'}_n)
   /\cosh(i\bar\eta^{'}_n b_1) \cos(-n \pi) , & \textrm{if} \ \beta_m \neq \bar\eta_n^{'}\nonumber.
\end{array} 
\right.   \\
\label{O}
   O_{mn}&=& \int_{-a_2}^{0} \left(\cosh(i\bar\alpha_{n}x)/\cosh(-i\bar\alpha_{n}b_{2})\right) \cos(n \pi)\Omega_{m}(x) dx\\
 &=& \left\{
\begin{array}{ll} 
a_2/\cosh(-i\bar\alpha_{n}b_{2})\cos(n \pi) , & \textrm{if} \  i \bar\alpha =\eta_m= 0,\\
(\sin(a_2\eta_m (1 - 1i)) (1 + 1i) + \sin(a_2 \eta_m (1 + 1i)) \\
(1 - 1i))/(4\eta_m)/\cosh(-i\bar\alpha_{n}b_{2})\cos(n \pi)
, & \textrm{if} \  i \bar\alpha =\eta_m\neq 0,\\
(\eta_m\cos(a_2\alpha_n)\sin(a_2\eta_m) - \alpha_n\cos(a_2\eta_m)\sin(a_2\alpha_n))\\
/(\eta_m^2 - \alpha_n^2)/\cosh(-i\bar\alpha_{n}b_{2})\cos(n \pi)
, & \textrm{if} \ i \bar\alpha \neq \eta_m. \nonumber
\end{array}
\right.\\
\label{E}
E_{mn} &=& \int_{-a_2}^{0} \Upsilon_{n}(x) \Omega_{m}(x) dx \\ &=&
    \left\{
\begin{array}{ll}  
a_2 , & \textrm{if} \ \eta^{'}_n=\eta_m=0, \\
a_2/2 + \sin(2 a_2\eta_m)/(4\eta_m) , & \textrm{if} \ \eta^{'}_n=\eta_m \neq0, \\
\sin(a_2 (\eta_m - \eta^{'}_n))/(2 (\eta_m - \eta^{'}_n)) + \sin(a_2 (\eta_m + \eta^{'}_n))\\
/(2 (\eta_m + \eta^{'}_n)),  & \textrm{if} \ \eta^{'}_{n} \neq \eta_{m}.\nonumber
\end{array} 
\right. \\
\label{R}
 R_{mn}&=& \int_{0}^{a_1} \overline\Gamma_{n}(y) \overline\Phi_m(y) dy\\
 &=& \left\{
\begin{array}{ll}  
a_1/2 - \sin(2 a_1\gamma_m)/(4\gamma_m)
, & \textrm{if}\ \gamma_m=\lambda_{n},\\
  -(\gamma_m\cos(a_1\gamma_m)\sin(a_1\lambda_n) - \lambda_n\cos(a_1\lambda_n)\\
  \sin(a_1\gamma_m) /(\gamma_m^2 - \lambda_n^2), & \textrm{if} \ \gamma_m \neq \lambda_{n}.
\end{array} 
\right. \nonumber\\
\label{Q}
 Q_{mn}&=& \int_{0}^{a_1} \left(\sinh(i\bar\zeta_{n}y)/\sinh(i\bar\zeta_{n}b_1) \right) \sin(-(2n+1)\pi/2) \overline\Phi_m(y) dy \\
     &=&\left\{
\begin{array}{ll}  
 -(\cos(a_1\gamma_m)\sinh(a_1\gamma_m) - \cosh(a_1\gamma_m)
 \sin(a_1\gamma_m))\\/(2\gamma_m)
 \sinh(i\bar\zeta_{n}b_1)\sin(-(2n+1)\pi/2)
, & \textrm{if}\ \gamma_m=i\bar\zeta_{n},\\
   -i(\gamma_m\cos(a_1\gamma_m)\sin(a_1 \bar\zeta_n)i - \bar\zeta_n\sin(a_1\gamma_m)
  \\ \cos(a_1\bar\zeta_n)i)
   /(\gamma_m^2 - \bar\zeta_n^2)  \sinh(i\bar\zeta_{n}b_1)\sin(-(2n+1)\pi/2), & \textrm{if} \ \gamma_m \neq i\bar\zeta_{n}.
\end{array} 
\right. \nonumber
\end{eqnarray}}

{\footnotesize
\begin{eqnarray}
\label{W}
   W_{mn}&=& \int_{-a_2}^{0} \left(\sinh(i\bar\lambda_{n}x)/\sinh(-i\bar\lambda_{n}b_{2})\right) \sin((2n+1)\pi/2) \overline\Omega_{m}(x)dx\\
 &=& \left\{
\begin{array}{ll}  
-(\cos(a_2\zeta_m)\sinh(a_2\zeta_m) - \cosh(a_2\zeta_m)\sin(a_2\zeta_m))/(2\zeta_m) \\
\sinh(-i\bar\lambda_{n}(b_{2}))\sin((2n+1)\pi/2), & \textrm{if} \  i \bar\lambda =\zeta^{'}_m\neq 0,\\
-(\bar\lambda_n\cos(a_2\bar\lambda_n)\sin(a_2\zeta_m)i - \zeta_m\cos(a_2\zeta_m)\sin(a_2\bar\lambda_n)i)\\
/(\bar\lambda_n^2 - \zeta_m^2)
\sinh(-i\bar\lambda_{n}(b_{2}))\sin((2n+1)\pi/2)
, & \textrm{if} \ i \bar\lambda \neq \zeta^{'}_m.
\end{array}
\right.\nonumber\\
\label{K}
K_{mn} &=& \int_{-a_2}^{0} \overline\Upsilon_{n}(x)\overline\Omega_{m}(x)dx \\ &=&
    \left\{
\begin{array}{ll}  
a_2/2 , & \textrm{if} \ \zeta_n=\zeta^{'}_m,\\
-(\zeta^{'}_m\cos(a_2\zeta^{'}_m)\sin(a_2\zeta_n)-\zeta_n\cos(a_2\zeta_n)\sin(a_2\zeta^{'}_m))\\
/(\zeta_m^{'2} - \zeta_n^2),  & \textrm{if} \ \zeta^{'}_{m} \neq \zeta_{n}.
\end{array} 
\right.  \nonumber   \\
\label{Vhat}
 V^{'}_{mn}&=& \int_{0}^{a_1} \left(\sinh(i\bar\eta^{'}_{n}y)/\sinh(i\bar\eta^{'}_{n}b_1) \right) \cos(-n\pi) \overline\Phi_{n}(y) dy \\
     &=&\left\{
\begin{array}{ll}  
 -(\cos(a_1\gamma_m)\sinh(a_1\gamma_m) - \cosh(a_1\gamma_m)
 \sin(a_1\gamma_m))/(2\gamma_m)\\
 \sinh(i\bar\eta^{'}_{n}b_1)
 \cos(-n\pi)
, & \textrm{if}\ \gamma_m=i\bar\eta^{'}_{n},\\
   -i(\gamma_m\cos(a_1\gamma_m)\sin(a_1 \bar\eta^{'}_n)i - \bar\eta^{'}_n\sin(a_1\gamma_m)
   \cos(a_1\bar\eta^{'}_n)i)\\/(\gamma_m^2 - \bar\eta^{2'}_n)
   \sinh(i\bar\eta^{'}_{n}b_1)\cos(-n\pi), & \textrm{if} \ \gamma_m \neq i\bar\eta^{'}_{n}.
\end{array} 
\right. \nonumber \\
\label{What}
 W^{'}_{mn}&=& \int_{-a_2}^{0} \left(\cosh(i\bar\lambda_{n}x)/\cosh(-i\bar\lambda_{n}b_{2})\right) \sin((2n+1)\pi/2) \cos\eta_{m}(x) dx\\
 &=& \left\{
\begin{array}{ll} 
a_2/\cosh(-i\bar\lambda_{n}b_{2})\sin((2n+1)\pi/2) , & \textrm{if} \  i \bar\lambda =\eta_m= 0,\\
(\sin(a_2\eta_m (1 - 1i)) (1 + 1i) + \sin(a_2 \eta_m (1 + 1i)) (1 - 1i))/(4\eta_m)\\
/\cosh(-i\bar\lambda_{n}b_{2})\sin((2n+1)\pi/2)
, & \textrm{if} \  i \bar\lambda =\eta_m\neq 0,\\
(\eta_m\cos(a_2\lambda_n)\sin(a_2\eta_m) - \lambda_n\cos(a_2\eta_m)\sin(a_2\lambda_n))/(\eta_m2 - \lambda_n2)
\\/\cosh(-i\bar\lambda_{n}b_{2})\sin((2n+1)\pi/2)
, & \textrm{if} \ i \bar\lambda\neq \eta_m.
\end{array}
\right.\nonumber\\
\label{Qhat}
   Q^{'}_{mn}&=& \int_{0}^{a_1}  \left(\cosh(i\bar  \zeta_n y)/\cosh(i\bar  \zeta_n b_1) \right)\sin ( -(2n+1)\pi/2) \cos \beta_{m} (y) dy \\
    &=& \left\{
\begin{array}{ll}  
 a_1/\cosh(i\bar\zeta_n b_1) \sin ( -(2n+1)\pi/2), & \textrm{if}\ \beta_m = \bar\zeta_n=0,\\
\sin(a_1\beta_m (1 - 1i)) (1 + 1i) + \sin(a_1\beta_m (1 + 1i)) (1 - 1i))\\/(4\beta_m)
/\cosh(i\bar\zeta_n b_1) \sin ( -(2n+1)\pi/2), & \textrm{if}\ \beta_m = \bar\zeta_n\neq 0,\\
   (\beta_m\cos(a_1\zeta_n)\sin(a_1\beta_m) - \bar\zeta_n\cos(a_1\beta_m)\sin(a_1\bar\zeta_n))\\
   /(\beta_m^2 - \bar\zeta_n)
   /\cosh(i\bar\zeta_n b_1) \sin ( -(2n+1)\pi/2) , & \textrm{if} \ \beta_m \neq \bar\zeta_n.
\end{array} 
\right.  \nonumber \\
\label{Ohat}
  O^{'}_{mn}&=& \int_{-a_2}^{0} \left(\sinh(i\bar\alpha_{n}x)/\sinh(-i\bar\alpha_{n}(b_{2}))\right) \cos(n \pi) \sin\zeta^{'}_{m}(x)dx\\
 &=& \left\{
\begin{array}{ll}  
-(\cos(a_2\zeta_m)\sinh(a_2\zeta_m) - \cosh(a_2\zeta_m)\sin(a_2\zeta_m))/(2\zeta_m) \\
\sinh(-i\bar\alpha_{n}(b_{2}))\cos(n \pi), & \textrm{if} \  i \bar\alpha =\zeta^{'}_m\neq 0,\\
-(\bar\alpha_n\cos(a_2\bar\alpha_n)\sin(a_2\zeta_m)i - \zeta_m\cos(a_2\zeta_m)\sin(a_2\bar\alpha_n)i)\\/(\bar\alpha_n^2 - \zeta_m^2)
\sinh(-i\bar\alpha_{n}(b_{2}))\cos(n \pi)
, & \textrm{if} \ i \bar\alpha \neq \zeta^{'}_m
\end{array}
\right.\nonumber
\end{eqnarray}}

\section{The result of the decomposition using symmetry }
\label{result}
\noindent Using the formula \eqref{11}, we can compute the complete solutions for the incident wave that is symmetric about $y=0$ to obtain  

{\footnotesize
\begin{eqnarray*}
     \phi^{\mathcal{N}}(x,y)&=& \left\{
\begin{array}{ll}  
e^{i\bar{\beta_{p}} (x+b_{2})} \cos\beta_{p}(y)+\displaystyle\sum_{n=0}^{\infty} \left(\frac{A^{\mathcal{NN}}_{n}+A^{\mathcal{D}\mathcal{N}}_{n}}{2}\right)e^{-i\bar{\beta_{n}} (x+b_{2})} \cos\beta_{n}(y), & \textrm{if} \ x\le -b_2;\\
\displaystyle\sum_{n=0}^{\infty} \left(\frac{B^{\mathcal{NN}}_{n}}{2}  \left( \frac{\cosh(i\bar \alpha_{n}x)}{\cosh(-i\bar \alpha_{n}b_2)}\right) +\frac{B^{\mathcal{D} \mathcal{N}}_{n}}{2}  \left(\frac{\sinh(i\bar \alpha_{n}x)}{\sinh(-i\bar \alpha_{n}b_2)}\right)\right) \cos \alpha_{n}(y), & \textrm{if} \ -b_2<x<0,\\
\displaystyle\sum_{n=0}^{\infty} \frac{C^{\mathcal{NN}}_{n}}{2}  \left(\frac{\cosh(i\bar \eta^{'}_n y)}{\cosh(i\bar  \eta^{'}_n b_1)} \right) \cos\eta^{'}_n (x)+\displaystyle\sum_{n=0}^{\infty} \frac{C^{\mathcal{D}\mathcal{N}}_{n}}{2} \left(\frac{\cosh(i\bar\zeta_{n}y)}{\cosh(i\bar\zeta_{n}b_1)} \right) \sin  \zeta_n (x), & \textrm{if} \ 0<x<b_2,\\
\displaystyle\sum_{n=0}^{\infty} \frac{D^{\mathcal{NN}}_n}{2} e^{i\bar{\eta_{n}}(y-b_1)}\cos \eta_{n} (x)+ \sum_{n=0}^{\infty} \frac{D^{\mathcal{D} \mathcal{N}}_n}{2}  e^{i\bar{\zeta^{'}_{n}}(y-b_1)}\sin \zeta^{'}_{n} (x), & \textrm{if} \  y\ge b_1,\\
\displaystyle\sum_{n=0}^{\infty} \frac{D^{\mathcal{NN}}_n}{2} e^{i\bar{\eta_{n}}(-y-b_1)}\cos \eta_{n} (x)+ \sum_{n=0}^{\infty} \frac{D^{\mathcal{D} \mathcal{N}}_n}{2}  e^{i\bar{\zeta^{'}_{n}}(-y-b_1)}\sin \zeta^{'}_{n} (x), & \textrm{if} \  y\le b_1,\\
\displaystyle\sum_{n=0}^{\infty} \left(\frac{A^{\mathcal{NN}}_{n}-A^{\mathcal{D}\mathcal{N}}_{n}}{2} \right) e^{-i\bar{\gamma_{n}} (-x+b_{2})} \sin\gamma_{n}(y), & \textrm{if} \ x\ge b_2.
\end{array} 
\right. 
\end{eqnarray*}}

\noindent By using the formula \eqref{22}, we can compute the solutions for the incident wave that is antisymmetric about $y=0$ to obtain

{\footnotesize
\begin{equation*}
     \phi^{\mathcal{D}}(x,y)= \left\{
\begin{array}{ll}  
e^{i\bar{\gamma_{p}} (x+b_{2})} \sin(\gamma_{p}y)+\displaystyle\sum_{n=0}^{\infty} \left(\frac{A^{\mathcal{DD}}_{n}+A^{\mathcal{N}\mathcal{D}}_{n}}{2}\right) e^{-i\bar{\gamma_{n}} (x+b_{2})} \displaystyle\sin(\gamma_{n}y), & \textrm{if} \ x<-b_2;\\
\displaystyle\sum_{n=0}^{\infty} \left(\frac{B^{\mathcal{DD}}_{n}}{2} \left( \frac{\cosh(i\bar \lambda_{n}x)}{\cosh(-i\bar \lambda_{n}b_2)}\right) +\displaystyle\frac{B^{\mathcal{N} \mathcal{D}}_{n}}{2} \left( \frac{\sinh(i\bar \lambda_{n}x)}{\sinh(-i\bar \lambda_{n}b_2)}\right)\right)\sin (\lambda_{n}y), & \textrm{if} \ -b_2<x<0,\\
\displaystyle\sum_{n=0}^{\infty} \frac{C^{\mathcal{DD}}_{n}}{2} \left( \frac{\sinh(i\bar  \eta^{'}_ny)}{\sinh(i\bar  \eta^{'}_n b_{1})}\right)  \cos(\eta^{'}_n x)+\displaystyle\sum_{n=0}^{\infty} \frac{C^{\mathcal{N} \mathcal{D}}_{n}}{2} \left( \frac{\sinh(i\bar  \zeta_{n}y)}{\sinh(i\bar  \zeta_{n}b_{1})}\right)  \sin ( \zeta_n x), & \textrm{if} \ 0<x<b_2,\\
\displaystyle\sum_{n=0}^{\infty} \frac{D^{\mathcal{DD}}_n}{2} e^{i\bar{\eta_{n}}(y-b_1)}\cos (\eta_{n}x)+\displaystyle \sum_{n=0}^{\infty} \frac{D^{\mathcal{N} \mathcal{D}}_n}{2}  e^{i\bar{\zeta^{'}_{n}}(y-b_1)}\sin (\zeta^{'}_{n}x), & \textrm{if} \  y>b_1, -a_2<x<a_2,\\
\displaystyle -\sum_{n=0}^{\infty} \frac{D^{\mathcal{DD}}_n}{2} e^{i\bar{\eta_{n}}(-y-b_1)}\cos(\eta_{n} x)+\displaystyle \sum_{n=0}^{\infty} \frac{D^{\mathcal{N} \mathcal{D}}_n}{2}  e^{i\bar{\zeta^{'}_{n}}(-y-b_1)}\sin (\zeta^{'}_{n} x), & \textrm{if} \  y<-b_1, -a_2<x<a_2,\\
\displaystyle \sum_{n=0}^{\infty} \left(\frac{A^{\mathcal{DD}}_{n}-A^{\mathcal{N} \mathcal{D}}_{n}}{2} \right) e^{-i\bar{\gamma_{n}} (-x+b_{2})} \sin(\gamma_{n}y), & \textrm{if} \ x>b_2.
\end{array} 
\right. 
\end{equation*}}

\bibliographystyle{unsrt}  
\bibliography{references}

\end{document}